\documentclass[twocolumn,prd,aps,floatfix]{revtex4}

\usepackage{cancel}
\usepackage{graphicx}
\usepackage{color}
\usepackage{amsmath}
\usepackage{amsfonts}
\usepackage{ulem}
\usepackage{comment}

\newcommand{\beq}{\begin{eqnarray}}
\newcommand{\eeq}{\end{eqnarray}}

\newcommand{\JPP}{J/\psi \textit{\rm -}\phi}
\newcommand{\JPR}{J/\psi \textit{\rm -}\rho}
\newcommand{\GammaL}{\Gamma^{(l)}}
\newcommand{\GammaS}{\Gamma^{(s)}}
\newcommand{\GammaO}{\Gamma_{\rm irrep}}

\usepackage{epstopdf}

\begin{document}

\title{L\"uscher's finite size method with twisted boundary conditions: \\
An application to the $J/\psi$-$\phi$ system to search for a narrow resonance}
\author{Sho Ozaki$^{1,2}$~\footnote{sho@rcnp.osaka-u.ac.jp} and
Shoichi Sasaki$^{3,1}$~\footnote{ssasaki@nucl.phys.tohoku.ac.jp}}

\affiliation{$^1$Theoretical Research Division, Nishina Center, RIKEN, Wako 351-0198, Japan}
\affiliation{$^2$Institute of Physics and Applied Physics, Yonsei University, Seoul 120-749, Korea}
\affiliation{$^3$Department of Physics, Tohoku University, Sendai 980-8578, Japan}

\date{\today}

\begin{abstract}

We investigate an application of twisted boundary conditions for the study of low-energy hadron-hadron interactions 
with L\"uscher's finite size method. This allows us to calculate 
the phase shifts for elastic scattering of two hadrons at any small value
of the scattering momentum even in a finite volume. We then can extract model-independent information
of low-energy scattering parameters, such as the scattering length, the effective range, and the effective volume from, 
the $S$-wave and $P$-wave scattering phase shifts through the effective range expansion.
This approach also enables us to examine the existence of near-threshold and narrow resonance states,
whose characteristics are observed in many of newly discovered charmonium-like $XYZ$ mesons.
As a simple example, we demonstrate our method for low-energy $J/\psi$-$\phi$ scatterings to search for 
$Y(4140)$ resonance using 2+1 flavor PACS-CS gauge configurations at the lightest pion mass, $m_{\pi}=156$ MeV.

\end{abstract}
\maketitle

\newpage

\section{Introduction}

In past several years, properties of hadronic interactions
have been extensively studied in lattice QCD 
simulations based on L\"uscher's finite size method, 
which is proposed as a general method for computing low-energy scattering 
phase shifts of two particles in finite volume~\cite{Luscher:1985dn, Luscher:1990ux}.
Various meson-meson, meson-baryon and baryon-baryon scattering 
lengths, which are relevant for describing low-energy scattering processes, 
have been successfully calculated in lattice QCD within this approach~\cite{Beane:2008dv}.

The original L\"uscher method, which is considered
for zero total momentum $\vec{P}=0$ of the two-particle system
in a symmetric box of size $L^3$ with periodic boundary conditions, 
has been developed in various ways~\cite{foot1}
. 
In particular, there are
several extensions of its formula for computing the scattering phase shifts
at more values of lower scattering momenta in a given volume.
From a practical viewpoint, accessible values of the phase shift on the lattice are limited due 
to discrete momenta, approximately, in units of $2\pi/L$.
To increase accessible momenta in a single volume, 
the formalism was extended to moving frames, where the 
scattering particles have nonzero total momentum $\vec{P}\neq 0$~\cite{Rummukainen:1995vs,Christ:2005gi,Kim:2005gf}, 
and also generalized in an asymmetric box of size $(\eta_1L)\times(\eta_2L)\times L$, 
where the degeneracy of low-lying modes in
three-dimensional momentum space can be resolved for $\eta_1, \eta_2 \neq 1$~\cite{Feng:2004ua, Li:2007ey}.
Alternatively, we notice that an idea of twisted boundary conditions
is quite useful for studying the hadron-hadron interaction at low energies
through L\"uscher's finite size method as originally proposed by Bedaque~\cite{Bedaque:2004kc}.

Since the twisted boundary conditions allow us to evaluate
scattering phase shifts of the two-particle system at any small value of the scattering momentum 
even in a finite box, detailed information of the low-energy interaction, which
is represented by the lower partial-wave phase shifts
at low energies, 
can be easily obtained through an extended L\"uscher formula for twisted boundary conditions.
In general, the quantity of $k^{2l+1}\cot \delta_l(k)$, where 
$\delta_l(k)$ and $k$ denote
the phase shift of the $l$th partial wave
and the absolute value of the scattering momentum ($k=|\vec{k}|$), 
can be expanded in a power series of the scattering momentum squared in the vicinity of the threshold
as
\begin{equation}
k^{2l+1} \cot \delta_l(k) = \frac{1}{a_l}
+\frac{1}{2}r_l k^2+{\cal O}(k^4),
\label{Eq:EffRangeExp}
\end{equation}
which is called the effective-range expansion~\cite{Newton:1982qc}. 
Model-independent information of the low-energy interaction is encoded in a small
set of parameters, {\it e.g.}, scattering length $a_0$ and effective range $r_0$ for an $S$ wave ($l=0$).
Therefore, we can determine such scattering parameters
through the $k^2$ dependence of the scattering phase shifts
near the threshold using the novel trick of twisted boundary conditions~\cite{Ozaki}.

Furthermore, we are able to apply this method to investigate newly observed narrow resonances in the heavy sector.
Recently, many charmonium- and bottomonium-like resonances, the so-called 
$XYZ$ \cite{Brambilla:2010cs, Godfrey:2008nc} and 
$Y_{b}Z_{b}$ \cite{Belle:2011aa, Collaboration:2011gja, Abe:2007tk, Chen:2008xia} resonances, are reported 
in several experiments.
Among them, some resonances are observed near two-particle thresholds, and their widths are quite narrow as compared to typical hadron resonances.
As a simple application for the new approach, we study a low-energy scattering of two mesons and search for a narrow resonance
near the threshold.
The $\JPP$ channel is considered to be an appropriate research target, since three narrow resonances have been 
reported in recent experiments, namely, $Y(4140)$ and $Y(4274)$ by the CDF 
Collaboration \cite{Aaltonen:2009tz, Aaltonen:2011at} and $X(4350)$ by the Belle Collaboration \cite{Shen:2009vs}.
Interestingly, these resonances seem to be relatively stable despite being 
above open charm thresholds, since the upper bound of their widths is less than $10$$-$$30$ MeV.
In particular, $Y(4140)$ is located close to the $J/\psi$-$\phi$ threshold. 
The $Y(4140)$ resonance 
was first reported by the CDF Collaboration with $3.8 \sigma$ statistics in 2009~\cite{Aaltonen:2009tz}.
The signal is observed in the invariant mass of the $J/\psi \phi$ pairs of the decay $B^{+} \to J/\psi \phi K^{+}$ from 
$p \bar{p}$ collisions at $\sqrt{s} = 1.96$ TeV.
A preliminary update of the CDF Collaboration analysis leads to the observation of the $Y(4140)$ 
with a statistical significance of more than $5 \sigma$~\cite{Aaltonen:2011at}. 
The mass and width are $4143.4^{+2.9}_{-3.0} \pm 0.6$ MeV and $15.3^{-10.4}_{-6.1} \pm 2.5$ MeV, respectively~\cite{Aaltonen:2011at}.
Although the observed mass is much higher than the $D\bar{D}$ threshold,  
the $Y(4140)$ resonance has a very narrow width.
This observation suggests that the
$Y(4140)$ scarcely couples to open charm channels such as $D\bar {D}$ and also $D_{s}^{+}D_{s}^{-}$.
On the other hand, the Belle and LHCb collaborations have not yet found the $Y(4140)$
in their experiments \cite{Shen:2009vs, Aaij:2012pz}.
Although the $Y(4140)$ might have very interesting properties, its existence is still controversial experimentally.
Our analysis of low-energy $J/\psi$-$\phi$ scatterings under the twisted boundary conditions could
give a new insight into the $Y(4140)$ resonance from first principles QCD.

The paper is organized as follows.
In Sec. II, after a brief introduction of the twisted boundary condition, 
we show the finite size formula with such particular boundary conditions.
Next, we apply our formula to the low-energy $J/\psi$-$\phi$ scattering in Sec. III, 
and then show our results in Sec. IV.
Finally, we summarize our study.

\section{Theoretical Framework}

\subsection{Boundary conditions}
Let us recall the ordinary periodic boundary condition in the spatial directions:
\beq
\Psi(\vec{x} + L \vec{\epsilon}_{i}, t ) = \Psi(\vec{x}, t)
\label{PBC}
\eeq
with the Cartesian unit vector $\vec{\epsilon}_{i}$ along the $i$ axis ($i= x,y,z$),
which provides the well-known quantization condition on three-momentum
for the noninteracting case:
$\vec{p} = (2\pi/L) \vec{n}$ with a vector of integers $\vec{n}\in \mathbb{Z}^{3}$.
A typical size of the smallest nonzero momentum under
periodic boundary conditions, {\it e.g.}, $|\vec{p}_{\rm min}|\approx 2\pi/L \sim 420$ MeV
for $L\sim 3$ fm and 250 MeV for $L\sim 5$ fm, is still too large to 
investigate the hadron-hadron scattering at low energies.

A novel idea, the twisted boundary condition, was proposed by Bedaque to
circumvent this issue~\cite{Bedaque:2004kc}. The twisted boundary condition
is a sort of generalization of the periodic boundary condition in the following way:
\beq
\Psi_\theta(\vec{x} + L \vec{\epsilon}_{i}, t) = e^{i \theta_{i}} \Psi_\theta(\vec{x}, t),
\label{TBC}
\eeq
where the angle variable $\theta$ is called the twist angle and $\Psi_{\theta}$ stands for either the elementary or composite field operator 
on which twisted boundary conditions are imposed.
Here $\theta_{i}=0$ corresponds to the ordinary periodic boundary condition as described above, 
while $\theta_{i} = \pi$ corresponds to the antiperiodic boundary condition.
Under twisted boundary conditions, the discretized momentum on the lattice should be
modified as 
\beq
\vec{p} = \frac{2\pi}{L}\left(\vec{n} + \frac{\vec{\theta}}{2\pi}\right)
\eeq
with a twist angle vector $\vec{\theta}=(\theta_x, \theta_y, \theta_z)$ for the free case.
In principle, we can have any value of momentum on the lattice through the variation of the twist angle, continuously.
In particular, in this paper, we want to emphasize that the lowest Fourier mode, $\vec{n}=(0,0,0)$, 
can still receive nonzero momentum, which can be set to an arbitrary small value even in a fixed spatial extent $L$, 
using this novel trick. 

Of particular interest is the case of {\it partially twisted} boundary conditions, where we impose twisted boundary conditions on 
the {\it valence} quark fields of specific flavor. Needless to say, if the twisted boundary conditions are imposed on 
the sea quark fields, huge computational cost is required to generate a new gauge ensemble for each twisted angle.
In addition, the authors of Ref. \cite{Sachrajda:2004mi} show that the finite volume correction due to the partially twisted conditions 
is exponentially suppressed as the spatial extent $L$ increases. Therefore, there is the practical advantage of the partially twisted boundary conditions.

For convenience of explanation,
we introduce new fields $q^{\prime}$, which are transformed
from the original fields $q_{\theta}$ where the twisted boundary conditions are
imposed as
\beq
q^{\prime}(\vec{x}, t) 
= e^{-i\vec{\theta}\cdot \vec{x}/L} q_{\theta}(\vec{x}, t).
\label{transf}
\eeq
These fields now turn out to satisfy the usual boundary conditions as
$q^{\prime}(\vec{x} + L\vec{\epsilon_{i}}) = q^{\prime}(\vec{x})$.
This redefinition of the quark fields affects only the hopping terms
that appear in lattice fermion actions. For Wilson-type fermions, 
the hopping terms are transformed as
\begin{widetext}
\beq
&&\sum_{\mu} \bar{q}_\theta(x) \left[(1-\gamma_{\mu})U_{x,\mu}\delta_{x+\mu, y}
+(1+\gamma_{\mu})U_{x-\mu, \mu}^{\dagger}\delta_{x-\mu, y} \right]q_\theta(y) \cr
&& =
\sum_{\mu}\bar{q}^{\prime}(x) \left[(1-\gamma_{\mu})U_{x,\mu}^{\prime}(\vec{\theta})\delta_{x+\mu, y}
+(1+\gamma_{\mu})U_{x-\mu, \mu}^{\prime \dagger}(\vec{\theta}) \delta_{x-\mu, y} \right]q^{\prime}(y)
\label{twisted_dirac} 
\eeq
\end{widetext}
with $U^{\prime}_{x,\mu}(\vec{\theta}) = e^{i\theta_{\mu}a/L} U_{x,\mu}$ and $\theta_{\mu} = (0, \vec{\theta})$.
Therefore, we can easily calculate the quark propagator subject to twisted boundary conditions through
the simple replacement of the link variables $\{U_{x,\mu}\}$ by $\{U^{\prime}_{x,\mu}(\vec{\theta})\}$ in the hopping terms. 

Using this technique, we can easily construct a hadronic interpolating operator for states with finite momenta.
We now consider the traditional meson interpolating operator as a local bilinear operator, 
${\cal O}_\Gamma(\vec{x},t)=\bar{q}_{f}^{\prime}(\vec{x}, t)\Gamma q_{f^{\prime}}^{\prime}(\vec{x},t)$,
where $f$, $f^{\prime}$ denote flavor indices with Dirac's gamma matrices $\Gamma$, as a simple example.
A simple summation over the spatial sites on this operator can be interpreted as 
the Fourier transformation to a momentum representation:
\beq
\sum_{\vec{x}}{\cal O}_\Gamma(\vec{x},t)
&=&\sum_{\vec{x}}\bar{q}_{\theta, f}(\vec{x}, t)\Gamma q_{\theta, f^{\prime}}(\vec{x},t)e^{-i(\vec{\theta}_{q_{f}}-\vec{\theta}_{\bar{q}_{f^{\prime}}})\cdot\vec{x}/L}\cr
&=& {\cal O}_{\Gamma}(\vec{p}, t),
\label{Eq:FFT_OP}
\eeq
where $\vec{p}=(\vec{\theta}_{q_f}-\vec{\theta}_{\bar{q}_{f^{\prime}}})/L$. Unless the same twisted boundary
conditions are imposed on both quark and antiquark operators, the resulting meson operator does receive finite momentum
because $\vec{\theta}_{q_f}\neq \vec{\theta}_{\bar{q}_{f^\prime}}$. This technique is widely used 
for flavorful mesons ($f\neq f^{\prime}$) in several lattice analyses~\cite{deDivitiis:2004kq, Boyle:2008yd, Kim:2010sd}.

\begin{figure}
\begin{minipage}{0.9\hsize}
\begin{center}
\includegraphics[width=0.9 \textwidth]{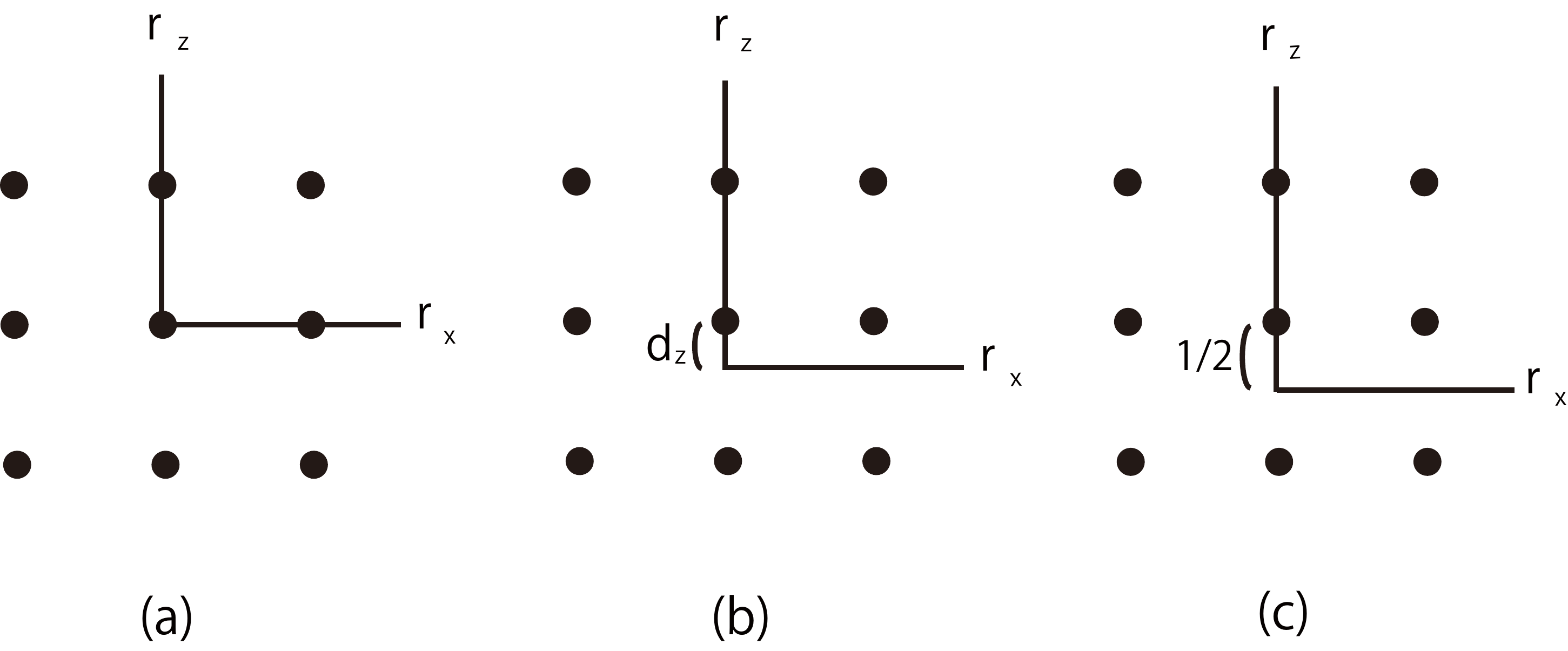}
\vskip -0.1in
\end{center}
\end{minipage}
\caption{The grid points in momentum space specified by a vector $\vec{r}$ are 
plotted in the $xz$ plane. For the case of a twisted angle vector 
$\vec{\theta} = (0, 0, \theta)$,  the grid points are shifted from the origin in the $z$ direction by $d_z=\theta/2\pi$.
The panels show the cases of (a) $\theta = 0$, (b) $0 < \theta < \pi$ and (c) $\theta = \pi$.
A center of symmetry is clearly lost in the middle panel. However, when $\theta = \pi$ (right panel), 
the origin is shifted at the midpoint of the grid interval, where the inversion center gets positioned.
}
\label{Fig:Qshift}
\end{figure}

\subsection{Finite size formula}
Let us consider the finite size formula under partially twisted boundary conditions.
In this study, we only consider the center-of-mass (CM) system with zero total momentum 
$\vec{P} = 0$, so that the CM system and the laboratory system coincide. Therefore, we do not have 
to consider the Lorentz boost from the laboratory frame to the CM frame in our approach, unlike
recent works that considered moving frames~\cite{Davoudi:2011md, Fu:2011xz, Leskovec:2012gb, Doring:2012eu, Gockeler:2012yj}.

First of all, the L\"uscher finite size formula provides a relation between finite volume corrections to the two-particle spectrum
and the physical scattering phase shift through a consideration of the relative two-particle wave function $\Psi(\vec{x}, t)$ 
in the CM frame. Since $\Psi(\vec{x}, t)$ defined in the CM frame 
should be independent of the relative time $t$, we simply omit the argument for the relative time hereafter.
Here the wave function $\Psi(\vec{x})$ is supposed to satisfy the Helmholtz equation in an outer range, 
where an interaction potential of finite range vanishes~\cite{Luscher:1990ux}. 
The Helmholtz equation is obtained with a set of wave numbers associated with discrete 
relative momenta $k^2$ in a finite box $L^3$:
\beq
(\Delta + k^2)\Psi(\vec{x})=0,
\eeq
where $k^2$ corresponds to the scattering momentum. Solutions of the Helmholtz equation are supposed to 
satisfy the twisted boundary conditions~(\ref{TBC}).
Here we introduce the Green function, which obeys the twisted boundary conditions:
\beq
G^{\vec{\theta}}(\vec{x}, k^2)=L^{-3}\sum_{\vec{p}\in\tilde{\Gamma}_{\vec{\theta}}}\frac{e^{i\vec{p}\cdot \vec{x}}}{\vec{p}^2 - k^2},
\eeq
where the momenta $\vec{p}$ are the elements of $\tilde{\Gamma}_{\vec{\theta}} = \{ \vec{p} | \vec{p} = \frac{2\pi}{L}\vec{n} + \frac{\vec{\theta}}{L}, \vec{n} \in \mathbb{Z}^{3} \}$. This function is of course a solution of the Helmholtz equation for $\vec{x}\neq0$ (mod $L$).
Further solutions may be obtained as its derivatives by using the harmonic polynomials, 
$\mathcal{Y}_{lm}(\vec{x}) = x^{l} Y_{lm}(\theta, \phi)$ defined with spherical coordinates, $\vec{x}=(x, \theta, \phi)$:
\beq
G_{lm}^{\vec{\theta}}(\vec{x}, k^2)={\cal Y}_{lm}(\vec{\nabla})G^{\vec{\theta}}(\vec{x}, k^2).
\eeq
These functions form a complete basis of the singular periodic solutions of the Helmholtz equation, which 
are supposed to have degree $\Lambda$.
The solutions can be represented by a linear combination of the functions $G_{lm}^{\vec{\theta}}(\vec{x}, k^2)$
with arbitrary coefficients $v_{lm}$:
\beq
\Psi(\vec{x})=\sum_{l=0}^{\Lambda}\sum_{m=-l}^{l}v_{lm}G_{lm}^{\vec{\theta}}(\vec{x}, k^2),
\eeq
which satisfies $\lim_{x\rightarrow 0}\left|x^{\Lambda+1}\Psi(\vec{x})\right|<\infty$.

Here it is worth mentioning that symmetry properties of $G^{\vec{\theta}}(\vec{x}, k^2)$
are not the same as the space group of the cubic lattice, namely the octahedral group $O_h$,
when $\vec{\theta}\neq 0$. 
This is simply because the finite momentum is induced by the twisted boundary conditions,
and therefore the symmetry of the reciprocal lattice is reduced to the subgroup of the full space group. 
Different partial-wave contributions are highly mixed in the expansion expression of $G_{lm}^{\vec{\theta}}(\vec{x}, k^2)$
in terms of spherical harmonics due to further reduction of the symmetry.

When we take $\vec{\theta} = 0$, the cubic symmetry is surely satisfied in both the space lattice and the reciprocal lattice.
The partial-wave mixing is maximally suppressed. 
In particular, even-$l$ and odd-$l$ waves do not mix with each other
because of a center of inversion symmetry in the cubic group.
The irreducible representation $A_{1g}$ of the cubic
symmetry contains partial waves of $l=0, 4, 6, 8, \cdot\cdot\cdot < \Lambda$ so that the wave function projected in the irrep $A_{1g}$
%
may still mix $l=0$ and $l=4$ waves at low energies~\cite{Luscher:1985dn, Luscher:1990ux}. 
As far as we consider the $S$-wave phase shift $\delta_0$ near the threshold,
the higher partial-wave ($l\ge4$) contributions are safely ignored. One may then choose 
an angular momentum cutoff as $\Lambda=0$ for the irrep $A_1$.
The original L\"uscher finite size formula for the $S$-wave phase shift $\delta_0$
is then given as the determinant of a $1\times1$ matrix~\cite{Luscher:1985dn, Luscher:1990ux} :
\beq
{\rm{cot}} \delta_{0}(k) = \frac{ 1 }{ \pi^{3/2} q } Z_{00}(1;q^{2}),
\label{Eq:OriginalFS}
\eeq
where $k$ and $q$ are the relative momentum of the two-particle system in the CM frame and 
its scaled momentum defined by $q^{2}=\left(\frac{ Lk }{ 2\pi }\right)^{2}$ with
the spatial extent $L$.  
Here the function $Z_{00}(s;q^{2})$ is the generalized zeta function, which is formally defined as
\beq
Z_{lm}(s;q^{2}) = \sum_{\vec{n} \in \mathbb{Z}^{3} } \frac{ \mathcal{Y}_{lm}(\vec{n}) }{ ( \vec{n}^{2} - q^{2} )^{s} }
\eeq
for $s>3/2$, and then an analytic continuation of the function is needed from 
the region $s>3/2$ to $s=1$~\cite{Luscher:1985dn, Luscher:1990ux}.

In the case of the twisted boundary condition with twist angles $0<|\theta| \le\pi$,
however, the cubic symmetry is broken into a subgroup symmetry.
For instance, when we take a twist angle vector $\vec{\theta} = (0, 0, \theta)$, which
is denoted symbolically by [001] hereafter, 
the grid points in momentum space are specified by a vector, $\vec{r} = \vec{n} + \vec{d}$ 
where $\vec{n} \in \mathbb{Z}^{3}$ and $\vec{d} = (0,0, \theta / 2\pi)$. 
In the case of $|\theta| \neq 0$, the mesh points are 
shifted in the $z$ direction by $d_{z} = \theta / 2\pi$, as schematically depicted in Fig~\ref{Fig:Qshift}.
Then, the symmetry of the reciprocal lattice, whose vector can be given 
by multiplying the vector $\vec{r}$ by a factor of $2\pi/L$,  is described by the little group $C_{4v}$.
Although the point group $C_{4v}$ does not have a center of inversion symmetry, 
it is recovered in the case of a special angle, $|\theta| = \pi$, where
the symmetry should be described by the $D_{4h}$ group as shown in Fig.~\ref{Fig:Qshift} (c).
Similarly, when $\vec{\theta} = (\theta, \theta, 0)$ (labeled by [110]), the symmetry is further reduced to point group 
$C_{2v}$ ($D_{2h}$) for $0<|\theta|<\pi$ ($|\theta|=\pi$), while
the symmetry is described by $C_{3v}$ ($D_{3d}$)
for $0<|\theta|<\pi$ ($|\theta|=\pi$) in the case of another twist angle vector $\vec{\theta} = (\theta, \theta, \theta)$, 
which is denoted symbolically by [111]. 

As mentioned previously, the point groups $C_{nv}$ for any finite $n$ do not have a center of inversion symmetry. 
As a consequence, even-$l$ and odd-$l$ partial waves would be mixed together under twisted boundary conditions
in the range of $0<|\theta|<\pi$. The possibility of such unwanted mixing is simply ignored in early exploratory works with twisted 
boundary conditions~\cite{{Kim:2010sd},{Kawanai:2010ru}}.
Recently, the same pathological issue is properly addressed in the derivation of 
Rummukainen-Gottlieb's formula (or, equivalently, L\"uscher's finite size formula in a moving frame) 
for the case of two particles with different masses~\cite{Fu:2011xz,Leskovec:2012gb, Doring:2012eu, Gockeler:2012yj}, 
where the symmetry of the system is reduced to the same little groups $C_{nv}$. 

The Rummukainen-Gottlieb finite size formula is known as a generalization of the L\"uscher formula for two-particle states 
calculated in the laboratory frame, where the total momentum of two particles is nonzero. 
The coordinate in the CM frame can be related to the original coordinate in the laboratory frame 
through the Lorentz transformation. In a moving frame approach, the original wave function, which is defined in the laboratory frame, 
can be Lorentz boosted into the appropriate CM frame with the Lorentz factor $\gamma$~\cite{foot2}
to derive the finite size formula, which provides a relation between finite volume corrections to the two-particle 
spectrum and the physical scattering phase shift.

In the original work~\cite{Rummukainen:1995vs}, where two degenerate states were considered, 
the factor simply introduces a negative sign in the boundary condition for the relative wave function. 
This indicates that the antiperiodic boundary condition, instead of the periodic boundary condition,
is imposed in the direction parallel to a nonvanishing component of the total momentum. 
However, for the case of two particles with different masses, the factor becomes a unit complex number, whose argument
depends on the mass difference~\cite{Davoudi:2011md, Fu:2011xz}. Consequently, the boundary condition for the 
relative wave function becomes identical to the twisted boundary condition, where the twist angle vector is parallel to 
the total momentum and the size of twist angles is associated with the mass difference of two-particles. 

\begin{table}[tb]
\begin{ruledtabular}
\begin{tabular}{cccc}
\hline
Twist angle & $(0, 0, \theta)$ & $(\theta, \theta, 0)$ & $(\theta, \theta, \theta)$ \\ 
Symmetry & $C_{4v}$ & $C_{2v}$ & $C_{3v}$ \\ 
Label & $[001]$ & $[110]$ & $[111]$ \\
\hline
${\cal M}^{\vec{\theta}}_{SS}$ & $w_{00}$ & $w_{00}$ & $w_{00}$ \\ 
${\cal M}^{\vec{\theta}}_{SP}$ & $ i\sqrt{3} w_{10}$ & $i\sqrt{6}w_{11}$ & $i 3 w_{10}$  \\ 
${\cal M}^{\vec{\theta}}_{PP}$ & $w_{00} + 2w_{20}$ & $ w_{00} - w_{20} -i \sqrt{6}w_{22}$ & $w_{00} - i2\sqrt{6} 
\omega_{22}$  \\
\hline
\end{tabular}
\caption{Matrix elements ${\cal M}^{\vec{\theta}}_{ab}$ for three different types of the twisted angle vector $\vec{\theta}$.
For simplicity, an irrelevant phase factor is omitted in the definition of ${\cal M}_{SP}^{\vec{\theta}}$.
The explicit expression of $w_{lm}$ is given in the text.
$w_{11}$ and $w_{22}$ are complex functions, while 
$w_{00}$, $w_{10}$ and $w_{20}$ are all real functions.
If special properties of ${\rm Re}(w_{11})={\rm Im}(w_{11})$ and ${\rm Re}(w_{22})=0$ are taken into account, 
$w_{11}$ and $w_{22}$ can be rewritten by a single real function as
$w_{11}=(1+i){\rm{Re}}(w_{11})$ and $w_{22}=i{\rm Im}(w_{22})$.}
\label{Table:MatrixDef}
\end{ruledtabular}
\end{table}

For the case of $\vec{P}=2\pi/L(0,0,1)$, Fu has derived the generalized Rummukainen-Gottlieb formula 
for the trivial $A_1$ sector of $C_{4v}$ as the determinant of a $2\times 2$ matrix, which is  composed 
of both $l=0$ and $l=1$ channels~\cite{Fu:2011xz}. Subsequently, the finite size formula for a moving frame is further generalized
for different irreducible representations of the reduced symmetry and also other types of nonvanishing total 
momenta: $\vec{P}=2\pi/L(1,1,0)$ and $2\pi/L(1,1,1)$~\cite{Leskovec:2012gb, Doring:2012eu, Gockeler:2012yj}.
We easily apply this knowledge to our twisted boundary cases in the CM frame, where the Lorentz boost factor 
is unity, $\gamma=1$.

The finite size formulas for $\vec{P}\neq 0$ kinematics~\cite{Fu:2011xz, Leskovec:2012gb, Doring:2012eu, Gockeler:2012yj} 
can be easily translated into the L\"uscher finite size formula
given under the partial twisted boundary conditions. 
Following the original expression for a moving frame~\cite{Fu:2011xz, Leskovec:2012gb, Doring:2012eu, Gockeler:2012yj}, 
when we take an assumption that $\delta_{l>1}=0$, the finite size formulas for the $A_{1}$ irrep are given as the following determinant 
%
%
%
\beq
\left|
\begin{array}{cc}
\cot \delta_0(k) - {\cal M}^{\vec{\theta}}_{SS}(q) &  {\cal M}^{\vec{\theta}}_{SP}(q) \cr
{\cal M}^{\vec{\theta}}_{SP}(q)^{\ast} & \cot \delta_1(k) - {\cal M}^{\vec{\theta}}_{PP}(q) \cr
\end{array}
\right|=0,
\label{Eq:MasterFS}
\eeq
which properly takes into account the mixing of $S$- and $P$-wave phase shifts
in one relation. The matrix elements ${\cal M}^{\vec{\theta}}_{SS}$, ${\cal M}^{\vec{\theta}}_{SP}$
and ${\cal M}^{\vec{\theta}}_{PP}$ that appeared in Eq.(\ref{Eq:MasterFS}) for three types of the twist angle vector 
are given in Table~\ref{Table:MatrixDef}, where we use the shorthand notation
\beq
w_{lm}(q) = \frac{1}{ \pi^{3/2} \sqrt{2l+1} q^{l+1} } Z_{lm}^{\vec{\theta}}(1;q^{2})^{\ast}.
\eeq
The generalized zeta function $Z_{lm}^{\vec{\theta}}(1;q^{2})$ with the twist angle vector is defined as
\beq
Z_{lm}^{\vec{\theta}}(s;q^{2}) = \sum_{\vec{r} \in \Gamma_{\vec{\theta}} } \frac{ \mathcal{Y}_{lm}(\vec{r}) }{ ( \vec{r}^{2} - q^{2} )^{s} }, 
\eeq
where the lattice grids $\vec{r}$ in the momentum space are the elements of $\Gamma_{\vec{\theta}} = \{ \vec{r} | \vec{r} = \vec{n} + \frac{\vec{\theta}}{2\pi}, \vec{n} \in \mathbb{Z}^{3} \}$.
For numerical evaluation of $Z_{lm}^{\vec{\theta}}(s;q^{2})$, we use a rapid convergent integral expression found in Appendix A of 
Ref.~\cite{Yamazaki:2004qb}.

The finite size formulas for three types of the twist angle vector are superficially quite 
different from each other, and this gives rise to some anxiety about
the possibility that there is no unique solution of the phase shift at given $k^2$.
However, we have numerically verified that when $|q| \ll 1/2$, 
${\cal M}^{[001]}_{ab}(q) \approx {\cal M}^{[110]}_{ab}(q) \approx {\cal M}^{[111]}_{ab}(q)$,
whereas when $1/2<|q| \ll \sqrt{2}/2$,  ${\cal M}^{[110]}_{ab}(q) \approx {\cal M}^{[111]}_{ab}(q)$.
Here we note that when we restrict ourselves to the region $|\theta|\le\pi$,
such a region of the twist angle gives the upper bound of the allowed 
kinematical region as $|q|\le \frac{1}{2}$ for the case of [001], $|q|\le \frac{\sqrt{2}}{2}$ for the case of [110], and
$|q|\le \frac{\sqrt{3}}{2}$ for the case of [111], respectively.
Therefore, three finite size formulas could approximately provide the identical values of the $S$-wave and $P$-wave
phase shifts obtained at the same scattering momenta within systematic uncertainties stemming from the reduction of 
the lattice rotational symmetry from the cubic symmetry $O_h$ 
to different little groups $C_{nv}$.

As mentioned previously, these modified finite size formulas show the mixing of $S$ and $P$ waves.
However, at $|\theta| = \pi$, the parity symmetry is restored, and  the $S$-wave and $P$-wave phase shifts 
are disentangled owing to the group theoretical constraint of ${\cal M}^{\vec{\theta}}_{SP}= 0$.
In this case, Eq.~(\ref{Eq:MasterFS}) gives a similar finite size formula to 
the original Rummukainen-Gottlieb one~\cite{Rummukainen:1995vs} without 
the Lorentz factor (since $\gamma=1$).
On the other hand, when $\vec{\theta}=(0,0,0)$, Eq.~(\ref{Eq:MasterFS}) properly reduces the original L\"uscher finite 
size formula (\ref{Eq:OriginalFS}) due to the fact that ${\cal M}^{\vec{\theta}}_{SP}=0$. 
Here we neglect higher partial-wave contributions above the $D$ wave ($l \ge 2$). Such contributions should be kinematically suppressed
as far as we consider low-energy hadron-hadron scattering near the threshold.

\section{Simulation details}

\subsection{Lattice setup}
We apply the L\"uscher finite size method that is generalized under the
partially twisted boundary conditions to explore $J/\psi$-$\phi$ scattering at low energies. 
A narrow resonance $Y(4140)$ that appeared in a $J/\psi$-$\phi$ decay mode has been reported by the CDF Collaboration.
If the $Y(4140)$ state really exists, we would directly observe a shape resonance in the $J/\psi$-$\phi$ scattering
using lattice QCD at the physical point.
We thus have performed dynamical lattice QCD simulations on a lattice $L^{3} \times T = 32^{3} \times 64$ 
with 2+1 flavor PACS-CS gauge configurations, where the simulated 
pion mass is closest to the physical point as $m_{\pi} = 156(7)$ MeV~\cite{Aoki:2008sm}.
Simulation parameters of PACS-CS gauge configurations are summarized in Table~\ref{Tab:Param}.
Our results are analyzed on all 198 gauge configurations, which are available through the International Lattice
Data Grid and the Japan Lattice Data Grid~\cite{ILDG}.

%
\begin{table*}
\begin{center}
\begin{ruledtabular}
\begin{tabular}{cccccc}
\hline
$\beta$ \ & \ $a$ (fm) \ &  $L^{3} \times T$ & \ $\sim La$ (fm) & $c_{\rm SW}$ \ &  \ Number of configs. \\ \hline
$1.9$  \ & $0.0907(13)$ \ & $32^{3} \times 64$ \ & $2.9$ \ &  $1.715$  \ &   $198$ \\
\hline
\end{tabular}
\caption{Parameters of $2+1$ flavor PACS-CS gauge configurations, generated using the Iwasaki gauge action
and Wilson clover fermions, at $m_{\pi} = 156(7)$ MeV and $m_K=554(2)$~\cite{Aoki:2008sm}.}
\label{Tab:Param}
\end{ruledtabular}
\end{center}
\end{table*}

We use nonperturbatively improved clover fermions 
for the strange quark ($s$) and a relativistic heavy quark (RHQ) action for the charm quark
($c$). The RHQ action is a variant of the Fermilab approach~\cite{ElKhadra:1996mp}, 
which can remove large discretization errors for heavy quarks. Parameters of clover fermions and the RHQ action used in this work 
are listed in Table~\ref{Tab:QuarkParam}. Although the simulated strange quarks are slightly off the physical point, the parameters 
are chosen to be equal to those of the strange sea quark used in gauge field generation. For the charm quark, adequate RHQ parameters defined in a Tsukuba-type action~\cite{Aoki:2001ra, Kayaba:2006cg} were calibrated to reproduce the experimental spin-averaged mass of a $1S$ charmonium state in Ref.~\cite{Kawanai:2011jt}.

\begin{table*}
\begin{center}
\begin{ruledtabular}
\begin{tabular}{ccccccccc}
\hline
Flavor & $\kappa$ & $\nu$ & $r_{s}$ & $c_{B}$ & $c_{E}$ & $M_V$ (GeV)  & Reference \\ \hline
Strange & $0.13640$ \ & $1.0$ \ & $1.0$ \ & $1.715$ \ & $1.715$ &  
1.0749(33)
& \cite{Aoki:2008sm} \\
Charm   & $0.10819$ \ & $1.2153$ \ & $1.2131$ \ & $2.0268$ \ & $1.7911$ & 3.0919(10)  & \cite{Kawanai:2011jt}\\
\hline
\end{tabular}
\caption{
Parameters of clover fermions (strange) and the RHQ action (charm) used in this work
and results of the vector meson masses. The strange quark mass is slightly off the physical point~\cite{Aoki:2008sm}.}
\label{Tab:QuarkParam}
\end{ruledtabular}
\end{center}
\end{table*}

\subsection{Interpolating operators}

In general, the Fourier transform of hadron interpolating operators
or the summation over all spatial points on the operators like Eq.~(\ref{Eq:FFT_OP}) {\it at the source time slice} 
is rather expensive from a computational point of view. The stochastic techniques are
often used for this purpose. 

We instead use the traditional gauge-fixed wall-source 
propagator and then compute either two-point functions for the $J/\psi$ and $\phi$ states or 
four-point functions of the $\JPP$ system using the wall-source operators.
As we will show later, the two-hadron operator constructed by the wall-source single-hadron operators is
automatically projected onto the trivial $A_1$ irrep of point groups $C_{nv}$ in the two-hadron system.

In the wall-source approach, the quark operator is summed over all spatial sites
at the source time slice, where Coulomb gauge fixing is performed. This can be interpreted as 
zero-momentum-state projection in the 
quark-level kinematics. Under the twisted boundary conditions, however, the wall source applied to the 
Dirac matrix inversion with the hopping term defined in Eq.~(\ref{twisted_dirac}) can be simply interpreted as the 
Fourier transformation to a momentum representation as 
\beq
\sum_{\vec x}q^{\prime}(\vec{x}, t_{\rm src})=\sum_{\vec x}q_\theta(\vec{x}, t_{\rm src})e^{-i\vec{\theta}\cdot \vec{x}/L}
\rightarrow q(\vec{p}, t_{\rm src})
\eeq
with momentum given by $\vec{p}=\vec{\theta}/L$. Thus, in this context, 
we directly construct the vector meson interpolating operator in three-dimensional momentum space from 
two wall sources with different twist angles $\vec{\theta}_{q_f}\neq \vec{\theta}_{\bar{q}_{f^\prime}}$ as
\beq
{\cal O}^W_\mu (\vec{p}, t_{\rm src})
=\sum_{\vec{x}}\bar{q}_{f}^{\prime}(\vec{x}, t_{\rm src})\gamma_\mu \sum_{\vec{y}}q_{f^\prime}^{\prime}(\vec{y}, t_{\rm src})
\eeq
with $\vec{p}=(\vec{\theta}_{q_{f}}-\vec{\theta}_{\bar{q}_{f^\prime}})/L$.
The superscript $W$ stands for wall. 

Of course, when the same twisted boundary conditions such that $\vec{\theta}_{q_f}=\vec{\theta}_{\bar{q}_{f^\prime}}$
are imposed on both quark and antiquark operators, the resulting meson operator does not receive finite momentum.
In the quarkonium case, such as $J/\psi$ and $\phi$, which are flavor-neutral mesons ($f=f^{\prime}$), 
$\vec{\theta}_{q_f}=\vec{\theta}_{\bar{q}_{f}}$ is a natural choice.
Therefore, for the quarkonium states, the twisted boundary condition trick does not seems to be applicable. 
However, we can practically choose $\vec{\theta}_{q_f}\neq \vec{\theta}_{\bar{q}_{f}}$ 
as {\it partially twisted boundary conditions} to construct the meson interpolating operator~\cite{Kawanai:2010ru}.
A price to pay is that, by construction, a disconnected diagram contribution in quarkonium two-point functions
is inevitably spoiled. In this context, the quarkonium states, which are flavor-neutral mesons, are fictitiously 
treated within this particular trick as if they were flavorful mesons consisting of two different flavors of the quark 
and antiquark with the same mass.

In reality, the contributions from the disconnected diagram in the vector channel are known to be negligibly 
small for strange and charm quarks in numerical simulations~\cite{{McNeile:2004wu},{deForcrand:2004ia}} 
in accordance with the mechanism of Okubo-Zweig-Iizuka suppression.
Then, even if the case of the usual periodic boundary condition, namely, $\vec{\theta}=0$,
is considered, the disconnected diagrams 
are often neglected in the $J/\psi$ and $\phi$ meson spectroscopy since their evaluations are computationally expensive.
Therefore, in this study, any disconnected diagram contribution
(such as self-annihilations of both $J/\psi$ and $\phi$) is not included in evaluating the two-point function 
of $J/\psi$ and $\phi$ mesons or the four-point functions of the $\JPP$ system in our simulations regardless of whether we use the $\vec{\theta}=0$ 
or $\vec{\theta}\neq 0$ case. For the twist angle, we choose $|\vec{\theta}_{q}|\neq 0$ and $|\vec{\theta}_{\bar{q}}|=0$ 
or vice versa for the interpolating operator construction.

\begin{table*}[htb]
\begin{ruledtabular}
\begin{tabular}{cccccccc}
\hline
&  $O_h$  &  $\downarrow C_{4v}$ & $\downarrow C_{2v}$ & $\downarrow C_{3v}$
\\ \hline
$\Gamma^{(s=0)}$ & $A_{1g}$& $A_{1}$ & $A_1$ & $A_1$\\ 
$\Gamma^{(s=1)}$ & $T_{1g}$ & $A_{2}\oplus E$ & $A_{2}\oplus B_1 \oplus B_2$ & $A_{2}\oplus E$\\ 
$\Gamma^{(s=2)}$ & $E_g \oplus T_{2g}$ & $(A_{1}\oplus  B_{1})\oplus(B_{2}\oplus E) $ 
& $(A_{1} \oplus B_{2})\oplus  (A_{1} \oplus A_{2} \oplus  B_{1})$& $E \oplus (A_{1}\oplus E)$ \\
\hline
\end{tabular}
\caption{Subduction of the irreps of $O_h$, which are obtained from the decomposition of the direct product representation 
$T_{1u}\otimes T_{1u}$ as the total spin of two vector particles, onto the little groups $C_{4v}$, $C_{2v}$ and $C_{3v}$. }
\label{Tab:Subdaction}
\end{ruledtabular}
\end{table*}

\begin{table*}[htb]
\begin{ruledtabular}
\begin{tabular}{cccccccc}
\hline
Twist angle vector& $(0,0,0)$ & $(0,0,\theta)$ & $(\theta,\theta,0)$ & $(\theta,\theta,\theta)$ & $(0,0,\pi)$ & $(\pi,\pi,0)$
&$(\pi,\pi,\pi)$ \\
Symmetry&  $O_h$  &  $C_{4v}$ & $C_{2v}$ & $C_{3v}$
& $D_{4h}$ & $D_{2h}$  & $D_{3d}$ 
\\ \hline
$\Gamma^{(l=0)}$ & $A_{1g}$& $A_{1}$ & $A_1$ & $A_1$
& $A_{1g}$ & $A_g$ & $A_{1g}$  \\ 
$\Gamma^{(l=1)}$ & $T_{1u}$ & $A_{1}\oplus E$ & $A_{1}\oplus B_1 \oplus B_2$ & $A_{1}\oplus E$
& $A_{2u}\oplus E_u$ & $B_{1u}\oplus B_{2u}\oplus B_{3u}$ & $A_{2u} \oplus E_{u}$\\ 
\hline
\end{tabular}
\caption{Classification of the decomposition of the representations $\Gamma^{(l=0)}$ and $\Gamma^{(l=1)}$
under each point group, which is associated with the choice of the twist angle vector $\vec{\theta}$. 
Point groups $O_h$, $D_{4h}$, $D_{2h}$ and $D_{3d}$ have
parity symmetry so that the trivial irrep $A_{1g}$ ($A_g$) does not contain the $l=1$ ($P$-wave) contribution. On the other hand, 
in the case of $C_{4v}$, $C_{2v}$ and $C_{3v}$, where the inversion center is lost, the
trivial irrep $A_1$ contains both the $l=0$ ($S$-wave) and $l=1$ ($P$-wave) contributions.}
\label{Tab:Classification}
\end{ruledtabular}
\end{table*}

At the sink we use a local bilinear operator as
\beq
{\cal O}^L_\mu(\vec{x}, t)=\bar{q}_f^{\prime}(\vec{x}, t)\gamma_\mu q^{\prime}_{f^{\prime}}(\vec{x},t),
\eeq
where $L$ stands for local and $f=f^\prime$ should be taken for the $J/\psi$ and
$\phi$ states. As described in Eq.~(\ref{Eq:FFT_OP}), under our choice of the partially twisted boundary conditions, 
a simple summation over the spatial sites on the sink operator automatically corresponds to the finite-momentum projection 
in terms of the Fourier transformation as ${\cal O}^L_{\mu}(\vec{p},t)=\sum_{\vec{x}}{\cal O}^L_{\mu}(\vec{x},t)$ with 
either $\vec{p}=\vec{\theta}_q/L$ or $-\vec{\theta}_{\bar{q}}/L$.
We then construct two-point functions for a single quarkonium state with finite momentum $\vec{p}$
using the wall-source and local sink operators:
\beq
G^{h}(\vec{p}, t, t_{\rm src})=\frac{1}{3}\sum_{i=1}^{3}\langle 
{\cal O}^{L, h}_{i}(\vec{p}, t){\cal O}_{i}^{W, h}(-\vec{p}, t_{\rm src})^{\dagger}
\rangle,
\eeq
where the superscript $h$ stands for either the $J/\psi$ or $\phi$ states.
We take an average over the spatial Lorentz indices $i$, which are related
to the polarization direction of the states so as to obtain
a possible reduction of statistical errors. Hereafter, we drop the label of 
both $W$ and $L$ on the single-hadron operators.

Next we consider two-hadron interpolating operators for the $\JPP$ system.
In order to consider the CM system, we assign momentum $\vec{p}$ to the $J/\psi$ operator and the opposite 
momentum $-\vec{p}$ to the $\phi$ operator through the partially twisted boundary conditions imposed on both 
the charm quark and the strange quark having the same twist angle but with the opposite sign.
The $\JPP$ interpolating operator for states 
with the relative momenta $\vec{p}$ of the $\JPP$ system are constructed from the product of the single $J/\psi$ 
and $\phi$ operators with opposite momentum:
\beq
Q_{ij}(\vec{p}, t) = O^{J/\psi}_{i} (\vec{p}, t)O^{\phi}_{j} (-\vec{p}, t+1),
\label{Eq:two-ope}
\eeq
where the subscripts $i$ and $j$ are spatial Lorentz indices.  To avoid the Fierz rearrangement of the 
$\JPP$ operator,  the $J/\psi$ and $\phi$ operators are displaced by one time slice. 
Then, the four-point function of the $\JPP$ system can be constructed from the above defined operators:
\beq
G_{ij; kl}^{\JPP}(\vec{p},t, t_{\rm src})=\langle Q_{ij}(\vec{p}, t)Q_{kl}(-\vec{p}, t_{\rm src})^{\dagger}\rangle.
\eeq

Here we assume that the $\JPP$ system can be treated as nonrelativistic. 
The two-hadron operator defined 
in Eq.~(\ref{Eq:two-ope}) can be described as the direct product of spin and momentum-space parts as 
\beq
Q_{ij}(\vec{p}, t)= S_{ij}\otimes Q(\vec{p}, t)
\eeq
within the nonrelativistic approximation, where the partial-wave contributions are built in the spin-independent part, 
$Q(\vec{p}, t)$. 
This treatment would be justified at least in the low-energy scattering region.
We then consider the partial wave ($l$) and spin ($s$) combination of the $\JPP$ system.
To identify the $\JPP$ state labeled by the angular momentum $l$ and spin $s$ in continuum, 
we have to identify the corresponding irreducible representations of
the discrete rotation group on the lattice. 

The $\JPP$ system has three different spin states: spin-0, spin-1 and spin-2 states. 
However, the $(2s+1)$-dimensional representation $\Gamma^{(s)}$ is, in general, reducible.
In the case of $\vec{\theta}=0$, namely, $\vec{p}=0$, the spin-1 vector particle at rest is assigned 
to the $T_{1u}$ irrep of the cubic group $O_h$~\cite{Moore:2005dw,Foley:2012wb}.
The total spin of two vector particles is given by the direct product representation $T_{1u}\otimes T_{1u}$, which is decomposed into
$A_{1g}\oplus E_{g} \oplus T_{1g} \oplus T_{2g}$, where $A_{1g}$ and $E_g$~\cite{foot3}
are one-dimensional and two-dimensional irreps respectively, while both $T_{1g}$ and $T_{2g}$ 
are three-dimensional irreps~\cite{Moore:2005dw,Foley:2012wb}.
The following decomposition rules are then obtained: 
\beq
\Gamma^{(s=0)} 
&=& A_{1g}, \nonumber \\
\Gamma^{(s=1)} 
&=& T_{1g},  \label{Eq:OhSpinDecomp} \\
\Gamma^{(s=2)} 
&=& T_{2g} \oplus  E_g. \nonumber
\eeq
For the cases of spin $0$ and spin $1$, there is a simple relationship between the cubic group 
representations and the continuum counterparts~\cite{Yokokawa:2006td}. Both spin-0 and spin-1 projections defined 
in the continuum theory remain valid for the corresponding irrep of the cubic group. 
We then obtain the respective four-point correlations for each irreducible representation of $O_h$ as
\begin{widetext}
\beq
G_{A_{1g}}^{\JPP}(t, t_{\rm src})&=&\frac{1}{3}\sum_{i=1}^3\sum_{j=1}^3G_{ii;jj}^{\JPP}(\vec{p}=0, t, t_{\rm src}),
\label{Eq:SpinDecompEqs_A1}
\\
G_{T_{1g}}^{\JPP}(t, t_{\rm src})&=&\frac{1}{6}\sum_{i=1}^3\sum_{j=1}^3\left\{
G_{ij;ij}^{\JPP}(\vec{p}=0, t, t_{\rm src})-G_{ij;ji}^{\JPP}(\vec{p}=0, t, t_{\rm src})
\right\},
\label{Eq:SpinDecompEqs_T1g}
\\
G_{T_{2g}}^{\JPP}(t, t_{\rm src})&=&\frac{1}{6}\sum_{i=1}^3\sum_{j=1}^3\left\{
G_{ij;ij}^{\JPP}(\vec{p}=0, t, t_{\rm src})+G_{ij;ji}^{\JPP}(\vec{p}=0, t, t_{\rm src})
-\frac{2}{3}G_{ii;ii}^{\JPP}(\vec{p}=0, t, t_{\rm src})
\right\},
\label{Eq:SpinDecompEqs_T2g}
 \\
G_{E_{g}}^{\JPP}(t, t_{\rm src})&=&\frac{1}{6}\sum_{i=1}^3\sum_{j=1}^3\left\{
G_{ii;ii}^{\JPP}(\vec{p}=0, t, t_{\rm src})-2G_{ii;jj}^{\JPP}(\vec{p}=0, t, t_{\rm src})
\right\}.
\label{Eq:SpinDecompEqs_Eg}
\eeq
\end{widetext}
As one can easily check, a linear combination of two four-point correlation functions, which are projected onto the $T_{2g}$ and $E_g$ irreps,
reproduces the spin-2 projection in the continuum theory~\cite{Yokokawa:2006td}. 
For the case of $\vec{\theta}=0$, where the cubic symmetry
is satisfied, the spin-0, spin-1 and spin-2 parts are orthogonal to each other. We then can easily determine the spin dependence of 
the $\JPP$ interaction through four types of four-point correlation functions defined in Eqs.~(\ref{Eq:SpinDecompEqs_A1})-(\ref{Eq:SpinDecompEqs_Eg}).

As we discussed in the previous section, for the cases of $\vec{\theta}\neq 0$, the symmetry of the reciprocal lattice 
becomes $C_{nv}$, which is the subgroup symmetry of the cubic group $O_h$, under the twisted boundary conditions.
The decomposition rules (\ref{Eq:OhSpinDecomp}) can be deduced by using the subduction of the irrep of $O_h$ to 
$C_{4v}$, $C_{2v}$ and $C_{3v}$ as summarized in Table~\ref{Tab:Subdaction}. 
As can be seen in the new decomposition rules, the spin-0 operator as the trivial irrep $A_1$ is inevitably mixed with
the spin-2 operator under the twisted boundary conditions with $\vec{\theta}\neq0$. 

To clarify this point, let us consider the case of the point group $C_{4v}$, where symmetry is realized in the 
case of a twist angle vector of $\vec{\theta}=(0,0,\theta)$. In this case, we may choose the 
corresponding spin-0, spin-1 and spin-2 operators
\beq
Q^{s=0}_{\JPP}(\vec{p}, t) &=& Q_{11}(\vec{p}, t)  + Q_{22}(\vec{p}, t)  + Q_{33}(\vec{p}, t), \cr
Q^{s=1}_{\JPP}(\vec{p}, t) &=& Q_{12}(\vec{p}, t)  - Q_{21}(\vec{p}, t), \cr
Q^{s=2}_{\JPP}(\vec{p}, t) &=& Q_{11}(\vec{p}, t)  + Q_{22}(\vec{p}, t)  - 2 Q_{33}(\vec{p}, t),
\nonumber
\eeq
which are guided by the knowledge in the continuum theory. The spin-1 operator 
transforms according to the $A_2$ irrep of $C_{4v}$, while both spin-0 and spin-2 operators transform according to 
the $A_1$ irrep. Although the spin-1 operator is certainly orthogonal to the spin-0 and spin-2 operators,
the latter ones are not orthogonal to each other in this case. Therefore, in practice, one may construct
the $2\times 2$ correlation matrix from $Q^{s=0}_{\JPP}(\vec{p}, t)$ and $Q^{s=2}_{\JPP}(\vec{p}, t)$
and then perform the variational method to disentangle the spin-0 and spin-2 contributions. 

Further reducing the symmetry from $C_{4v}$ to $C_{2v}$, both the spin-1 and spin-2 parts are involved in the same irrep ($A_2$).
Therefore, in this case, the disentanglement between the spin-1 and spin-2 contributions is also required.
Although the energy shift measured in the lower spin channel is slightly larger than the higher spin channel, 
we do not find the appreciable spin dependence in the $\JPP$ system through the calculation with $\vec{\theta}=0$,
where the spin-0, spin-1 and spin-2 contributions are clearly disentangled as discussed previously.
A similar finding in a quenched study of the $\JPR$ interaction with the usual periodic boundary conditions
was reported in Ref.~\cite{Yokokawa:2006td}. This finding indicates that the spin-independent part of the $\JPP$ interaction
dominates at least at low energies. Therefore, in this study, we will not resolve each spin contribution in simulations 
under the twisted boundary conditions, where the spin projection is somewhat complicated. Rather, we would like to 
resolve the partial-wave mixing between even-$l$ and odd-$l$ waves due to the twisted boundary conditions.

Our aim in this study is to demonstrate the feasibility of our proposed approach, where both the $S$-wave 
and $P$-wave phase shifts are extracted through the generalized L\"uscher finite size formula (\ref{Eq:MasterFS}),
as we will explain later. We focus on the spin-independent part of the $\JPP$ system 
for this purpose and thus use the spherical averaged (sav) $\JPP$ operator for all types of the twisted angle vector $\vec{\theta}$:
\beq
Q_{\JPP}^{\rm sav}(\vec{p}, t) = \frac{1}{3}\sum_{i=1}^3Q_{ii}(\vec{p}, t),
\eeq
which transforms with respect to the rotation of the spin direction 
according to the trivial irreducible representation of any point groups considered here.
Then, the resulting four-point function turns out to be identical to the one defined in Eq.~(\ref{Eq:SpinDecompEqs_A1})
for the irrep $A_{1g}$ of $O_h$.

Next we discuss the partial-wave contributions on the spherical averaged $\JPP$ operator. 
The twisted boundary conditions make the symmetry of the reciprocal lattice break down to
a subgroup symmetry of the cubic group $O_{h}$.
Therefore, the $(2l+1)$-dimensional representation $\GammaL$ becomes reducible
and should be decomposed into the irreducible representations of the group $C_{nv}$.
For instance, in the case of $\vec{\theta}=(0,0,\theta)$, 
there are four one-dimensional irreducible representations ($A_1$, $A_2$, $B_1$, $B_2$) and one
two-dimensional irreducible representation ($E$). For the angular momenta $l\le 2$, 
the resulting decompositions are given \cite{Fu:2011xz, Leskovec:2012gb, Doring:2012eu,Gockeler:2012yj} as
\beq
\Gamma^{(l=0)} 
&=& A_{1}, \nonumber \\
\Gamma^{(l=1)} 
&=& A_{1} \oplus E, 
\label{Eq:AMDecomp}
\\
\Gamma^{(l=2)} 
&=& A_{1} \oplus B_{1} \oplus B_{2} \oplus E.
\nonumber
\eeq
This clearly indicates that the mixing between 
even-$l$ and odd-$l$ wave contributions is not prohibited in some irreps.
In this paper, we take advantage of this mixing in order to
extract both $S$-wave and $P$-wave phase shifts simultaneously from the trivial $A_1$ irrep of the two-hadron operator, 
which certainly contains both $l=0$ and $l=1$ contributions. 
There is nothing to change for either $C_{2v}$ or $C_{3v}$, where the trivial irrep $A_1$ contains both $S$-wave and $P$-wave contributions, like in $C_{4v}$, as summarized in Table~\ref{Tab:Classification}.
As for the higher partial-wave contributions, we simply assume that they can be neglected.

Here we note that the decomposition for the partial wave $\GammaL$ is almost similar to 
the case of the spin $\GammaS$ found in Table~\ref{Tab:Subdaction},
except that the $A_1$ irrep, instead of the $A_2$ irrep, appears in the second line. The rules are
determined by the subduction of the irrep of the $O_h$ group to the $C_{4v}$ group.
The angular momentum $l=1$ is assigned to a vectorlike irreducible representation of $O_h$, 
which should be $T_{1u}$.
The $T_{1u}$ irrep of $O_h$ is subdued to the $A_1$ irrep of $C_{4v}$~\cite{Foley:2012wb} that appears 
in the second line of Eq.~(\ref{Eq:AMDecomp}), while the subduction of the $T_{1g}$ irrep is considered in 
Table~\ref{Tab:Subdaction} for the total spin of two vector particles.

In general, the operator that is projected onto some irrep $\GammaO$ of the group $G$ is simply given by
\beq
\hat{P}_{\GammaO}Q(\vec{p}, t)
&=& \frac{1}{g_G} \sum_{R \in G} \chi_{\GammaO}^{*}(R) \hat{\cal R} Q( \vec{p}, t) \nonumber \\
&=& \frac{1}{g_G} \sum_{R \in G} \chi_{\GammaO}^{*}(R)  Q( R \vec{p}, t), \nonumber
\eeq
where $g_G$ is the number of elements $R$ of the group $G$, while
$\chi_{\GammaO}(R)$ is a character of $R \in G$ for the irrep $\GammaO$.
Here we consider that $G$ is given by the groups $C_{nv}$. 
In this case, the momentum $\vec{p}$ is invariant by any transformation $R \in G$, namely, 
$R \vec{p} = \vec{p}$. Therefore, we get
\beq
\hat{P}_{\GammaO}Q(\vec{p}, t)=\frac{1}{g_G} \left(\sum_{R \in G} \chi_{\GammaO}^{*}(R)\right) Q(\vec{p}, t),
\label{Eq:ProjOP}
\eeq
where $g_G=\sum_{R \in G}1$.
Consequently, we observe that the right-hand side (r.h.s.) of Eq.~(\ref{Eq:ProjOP}) 
vanishes due to $\sum_{R \in G} \chi_{\GammaO}^{*}(R)=0$,
except for the case of the trivial irrep $A_1$, where $\sum_{R \in G} \chi_{A_1}^{*}(R)=g_G \neq 0$ 
is satisfied. This means that the operator $Q(\vec{p}, t)$ is already irreducible and should transform according to 
the trivial irrep of the groups $C_{nv}$~\cite{foot4}
. 
This is a consequence of the wall-source propagator, where the quark operators 
are summed over all spatial sites at the source time slice.
Thus, we can simply use the finite size formula of the $A_{1}$ sector given in Eq.~(\ref{Eq:MasterFS})
since the two-hadron operator used in our simulations is constructed by the wall-source single-hadron operators.

\section{Numerical results}

We can read off the single-hadron energy from the two-point functions under the partial twisted boundary conditions:
\beq
G^{h}(\vec{p}, t, t_{\rm src})\xrightarrow[t \gg t_{\rm src} ]{} e^{-E_h (t-t_{\rm src})},
\eeq
where $E_{h} = \sqrt{ \vec{p}^{ \ 2} + M_{h}^{2} }$, with the rest mass $M_h$, which can be determined with
the ordinary periodic boundary conditions ($\vec{\theta}=0$). It is worth mentioning that Dirichlet boundary conditions 
are imposed for all quarks in the time direction in order to avoid wrapround effects, which are very cumbersome 
in systems of more than two hadrons. Each hadron mass is then obtained by fitting the corresponding two-point 
correlation function with a single exponential form. The $\phi$ and $J/\psi$ masses are tabulated in Table~\ref{Tab:QuarkParam}.

\begin{figure}[t]
\begin{minipage}{\hsize}
\begin{center}
\includegraphics[width=0.9 \textwidth]{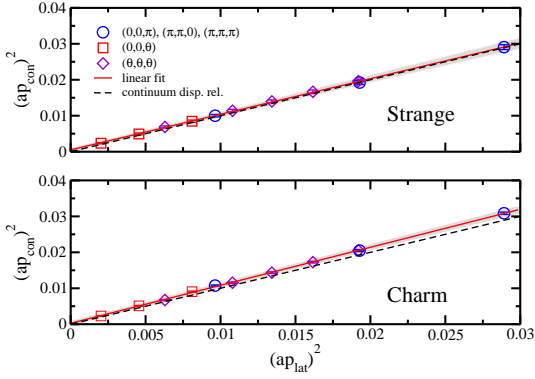}
\end{center}
\end{minipage}
\caption{Check of the dispersion relation for the $\phi$ (upper panel) and $J/\psi$ (lower panel) states.
The vertical axis shows the momentum squared defined through the relativistic continuum dispersion relation
as $\vec{p}^2_{\rm con}=E_h^2-M_h^2$ for $h=\phi$ and $J/\psi$, while the horizontal axis is the momentum 
squared defined by the given twist angles as $\vec{p}_{\rm lat}^2=\vec{\theta}^2/L^2$. 
By the linear fit (solid line) to data points calculated with various twist angles, the effective speed of
light is obtained as $c_{\rm eff}^2=0.990(44)$ for the strange sector and $c_{\rm eff}^2=1.057(19)$ 
for the charm sector. For comparison, the continuum dispersion relation is denoted as the dotted line in each panel.
}
\label{Fig:DispRel}
\end{figure}

Here we assume that each state ($h=J/\psi$ and $\phi$) 
on the lattice has the relativistic continuum dispersion relation, which is indeed enough to describe our data
as shown in Fig.~\ref{Fig:DispRel}. 
On the other hand, the scattering momentum $k$ of the $\JPP$ two-particle 
system is defined through 
\beq
W=\sqrt{ \vec{k}^{ \ 2} + M_{J/\psi}^{2} } + \sqrt{ \vec{k}^{ \ 2} + M_{\phi}^2 },
\eeq
where $W$ denotes the total energy of the $\JPP$ system in the CM frame. The CM energy can be determined from the large-$t$ behavior of four-point functions
\beq
G_{\JPP}(\vec{p}, t, t_{\rm src})\xrightarrow[t \gg t_{\rm src} ]{} e^{-W (t-t_{\rm src})}.
\label{Eq:FourPointFunc}
\eeq
Next let us define the two-particle energy $E$ measured from the threshold as
\beq
E = W - (M_{J/\psi} + M_{\phi}).
\label{relative_E}
\eeq
The scattering momentum then can be represented in terms of $E$, 
\beq
k 
&=&  \frac{ \sqrt{ E ( E + 2M )( E + 2M_{J/\psi} )( E + 2M_{\phi} ) } }{ 2 (E+M) }
\label{relati_momenta}
\eeq
with the sum of the rest masses $M = M_{J/\psi} + M_{\phi}$ and the individual masses.
Therefore, the precise determination of $E$ is a key point of the finite size analysis.
Instead of trying to directly measure $E$, we first measure the energy shift $\delta E$ 
from the ratio of four-point and two-point functions with the partially twisted boundary conditions
\beq
&&R_{\JPP}(\vec{p}, t) 
= \frac{ G_{\JPP}(\vec{p}, t,t_{\rm src}) }{ G^{J/\psi}(\vec{p}, t, t_{\rm src}) G^{\phi}(\vec{p}, t+1, t_{\rm src}+1) } \nonumber \\
&&\xrightarrow[t \gg t_{\rm src}]{}  \ \ 
 \exp \{- \delta E (t-t_{\rm src})\}, 
\label{Eq:RatioR}
\eeq
which reduces the statistical fluctuation due to a strong correlation between the denominator and numerator in the ratio.
We then obtain the energy $E$ from the following relation:
\beq
E = \delta E + \epsilon_{J/\psi} + \epsilon_{\phi},
\label{relative_E}
\eeq
where $\epsilon_{h} = E_{h} - M_{h}$ for $h=J/\psi$ and $\phi$, which can be evaluated
with the individual masses through the continuum-like dispersion relation. This procedure
for the determination of $E$ is similar to what was proposed in Ref.~\cite{Kim:2010sd}.

%
%
%
%

\begin{figure}[t]
\begin{minipage}{\hsize}
\begin{center}
\includegraphics[width=0.9\textwidth]{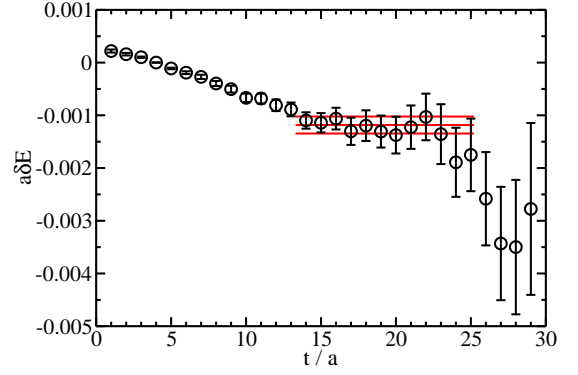}
\end{center}
\end{minipage}
\caption{The effective energy shift $\delta E$ in lattice units as a function of the time slice $t$,
as in the case of a twist angle $\vec{\theta} = (0, 0, \theta)$ with $\theta = 0.72$ (rad).}
\label{Fig:EffEng}
\end{figure}

In Fig.~\ref{Fig:EffEng}, 
we plot the effective energy shift defined by 
\beq
\delta E_{\rm eff}(\vec{p}, t)=\ln \frac{R_{\JPP}(\vec{p}, t) }{R_{\JPP}(\vec{p}, t+1) },
\eeq
which should show a plateau for large Euclidean time ($t\gg t_{\rm src}$), for the case of a twist angle $\vec{\theta} = (0,0, \theta)$ 
with $\theta = 0.72$ (rad) as a typical example. The quoted errors are estimated by a single elimination jackknife method. 
An important observation is that the negative energy shift found in Fig.~\ref{Fig:EffEng}
implies the presence of an attractive interaction between the $J/\psi$ and $\phi$ states. 

We find appropriate temporal windows, where the ground state dominance is satisfied, for fitting the ratio 
function defined in Eq.~(\ref{Eq:RatioR}). 
Three horizontal solid lines represent the fit result with its 1 standard deviation 
obtained by a covariant single exponential fit over the range of $14\le t/a \le 25$, where
two-point functions of the individual hadrons are also marginally dominated by the ground state of each hadron. 
We measure the energy shifts $\delta E$ for several twist angles and then observe that 
$\delta E$ measured in the $J/\psi \textit{\rm -}\phi$ system is almost unchanged with the variation of 
the CM energy $W$.

Once we obtain the scattering momentum $k$ of the two-particle system, we can calculate the scattering phase 
shifts through a set of three finite size formulas (\ref{Eq:MasterFS})
with three types of partially twisted 
boundary conditions, namely, $[001]$, $[110]$ and $[111]$. 
These formulas contain two unknowns, namely the $S$-wave and $P$-wave phase shifts. 
In order to obtain both phase shifts at various scattering momenta, we propose the following strategy.

\begin{figure}[h]
\begin{minipage}{\hsize}
\begin{center}
\includegraphics[width=0.9 \textwidth]{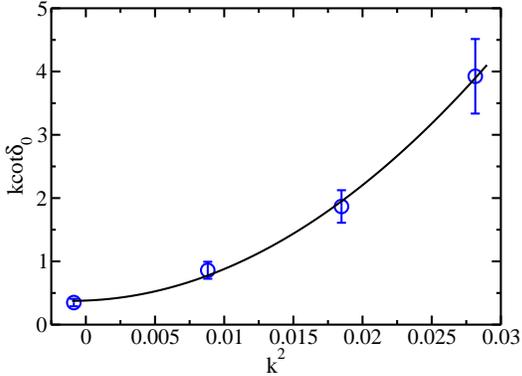}
\end{center}
\end{minipage}
\caption{
$k{\rm{cot}}\delta_{0}$ as a function of $k^2$ in lattice units. Open circles are calculated with twist angles, 
$\vec{\theta} = (0,0,0)$, $(0,0,\pi)$, $(\pi, \pi, 0)$ and $(\pi, \pi, \pi)$. A solid curve represents an interpolation of four data points
using the effective range expansion up to the order of $k^4$.
}
\label{Fig:K1cotD0_S}
\end{figure}

First, we calculate the $S$-wave phase shift $\delta_0(k)$ in the case of four special twist angles, $\vec{\theta} = (0,0,0)$, $(0,0,\pi)$, $(\pi, \pi, 0)$ and $(\pi, \pi, \pi)$, where the parity symmetry remains preserved.
Since there is no unwanted mixing between even-$l$ and odd-$l$ wave
contributions, the finite size formula reduces to either the L\"uscher formula or the Rummukainen-Gottlieb$-$type formula without the Lorentz factor
(since $\gamma=1$). 
In Fig.~\ref{Fig:K1cotD0_S}, we plot results of $k{\rm{cot}}\delta_{0}$, which are obtained from the data calculated with specific twist angles, $\vec{\theta} = (0,0,0)$, $(0,0,\pi)$, $(\pi, \pi, 0)$ and $(\pi, \pi, \pi)$ as a function of $k^2$ in lattice units. 
We observe that $k{\rm{cot}}\delta_{0}$ monotonically increases as $k^2$ increases and
there is no nonanalytic behavior in the range covered here. Therefore, we simply interpolate 
the four data points by the quadratic function of $k^2$ using the jackknife method.

Next, using the above information of the $S$-wave phase shift $\delta_0$,
we evaluate ${\rm{cot}}\delta_{1}$ from the data of the scattering momentum $k^2$ calculated 
with a twist angle vector, $\vec{\theta} = (\theta, \theta, \theta)$ through the formula 
\beq
{\rm{cot}}\delta_{1}(k) &=& {\cal M}^{[111]}_{PP}(q) + \frac{ | {\cal M}^{[111]}_{SP}(q) |^{2} }{ {\rm{cot}}\delta_{0}(k) - {\cal M}^{[111]}_{SS}(q) }
\cr
&=&w_{00}(q) + 2\sqrt{6}{\rm Im}\left\{w_{22}(q)\right\}+\frac{9w_{10}^2(q)}{\cot \delta_0(k)-w_{00}(q)}.
\label{Eq:FVFormula_111}\nonumber\\
\eeq
The results obtained for the $P$-wave phase shift $\delta_1$ are plotted as $k^3{\rm{cot}}\delta_{1}$ vs $k^{2}$ in Fig. 5. 
Although a slight but monotonic increase is observed, the results of $k^3{\rm{cot}}\delta_{1}$ show only very 
mild $k^2$ dependence. This observation suggests that the effective-range expansion is valid in this $k^2$ region.

%
%
\begin{figure}[h]
\begin{minipage}{\hsize}
\begin{center}
\includegraphics[width=0.9 \textwidth]{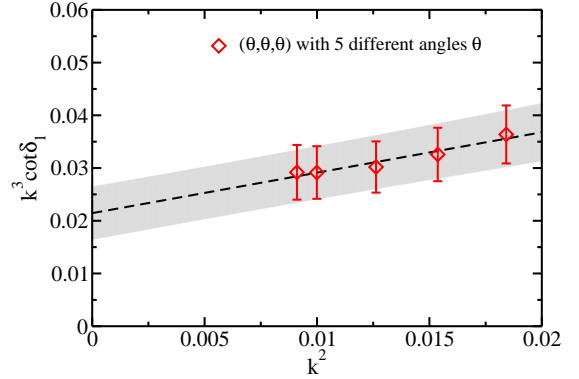}
\end{center}
\end{minipage}
\caption{$k^{3}{\rm{cot}}\delta_{1}$ as a function of $k^2$ in lattice units.
Open diamonds are calculated with a twist angle vector $\vec{\theta} = (\theta, \theta, \theta)$
with five different angles $\theta$. For evaluation of ${\rm{cot}}\delta_{1}$,
Eq.~(\ref{Eq:FVFormula_111}) is used with an input of ${\rm{cot}}\delta_{0}$, which
was determined in Fig.~\ref{Fig:K1cotD0_S}.}
\label{Fig:K3cotD1}
\end{figure}
%

In this work, we use five different angles, $\theta = 1.84$, 1.92, 2.14, 2.35 and 2.56 (rad),  
to choose $k^2$ points that range from $\left(\frac{\pi}{L}\right)^2$ to $2\left(\frac{\pi}{L}\right)^2$ in lattice units.
This is because when the value of $\theta$ becomes very small, the resulting $k^2$ is close to 
the threshold, $k^2\sim 0$. As we will discuss later, the $P$-wave mixing contributions to the finite size formula (\ref{Eq:MasterFS})
become negligible in the region of $k^2 < \left(\frac{\pi}{L}\right)^2$
so that the term of ${\rm{cot}}\delta_{0} - w_{00}$ is almost vanishing. 
This gives rise to numerical difficulties 
in this region for extracting $\delta_1$ from Eq.~(\ref{Eq:FVFormula_111}), where the second term 
on the right-hand side is very sensitive to how precisely the scattering momentum $k^2$ is determined.

To avoid the above numerical problem, we evaluate $k^{3} {\rm{cot}}\delta_{1}$ up to $k^2 \sim 0.008$ from $0.02$ in lattice units,
where the term of ${\rm{cot}}\delta_{0} - w_{00}$ is determined precisely enough, and we then extrapolate the value 
of $k^{3}{\rm{cot}}\delta_{1}$ to the region $k^{2} \sim 0$ by the polynomial function of $k^2$, which corresponds to the 
effective-range expansion~\cite{foot5}
. Thanks to very mild $k^2$ dependence, the linear function of $k^2$ is enough
to extrapolate the data of $k^{3} {\rm{cot}}\delta_{1}$ from the current set of points toward the low $k^2$ region, since the 
higher-order terms in the $k^2$ expansion should be suppressed near the threshold.

We thus obtain one of  the threshold parameters for the $P$ wave, the scattering volume $a_1$,
which is defined as an inverse of $k^3{\rm{cot}}\delta_{1}$ determined at $k=0$:
\beq
a_{1} =\left. \frac{ {\rm{tan}}\delta_{1} }{ k^{3} } \right|_{k=0} = 0.0348 \pm 0.0081 \ \ [{\rm{fm}}^{3}]
\eeq
through the linear fit on the data of $k^{3} {\rm{cot}}\delta_{1}$. The result of $a_1$ indicates that 
the $P$-wave interaction of the $J/\psi$-$\phi$ system is weakly attractive.

%
%
\begin{figure}[t]
\begin{minipage}{\hsize}
\begin{center}
\includegraphics[width=0.9 \textwidth]{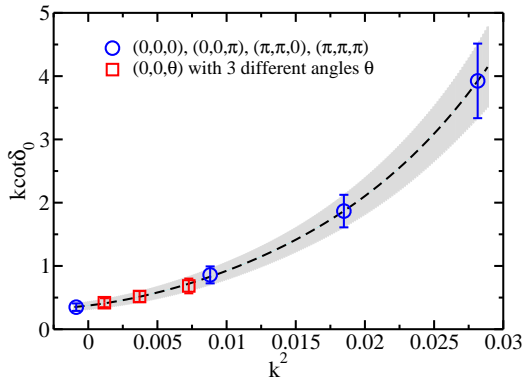}
\end{center}
\end{minipage}
\caption{$k{\rm{cot}}\delta_{0}$ as a function of $k^{2}$ in lattice units.
with $\vec{\theta} = (0, 0, \theta)$ data. The same data that appeared in Fig.~\ref{Fig:K1cotD0_S}  
are shown as open circles, while open squares stand for new data
calculated with another twist angle vector $\vec{\theta} = (0,0,\theta)$ with three different twist angles. 
For the evaluation of ${\rm{cot}}\delta_{0}$, 
Eq.~(\ref{Eq:FVFormula_100}) is used with an input of ${\rm{cot}}\delta_{1}$, which was
determined in Fig.\ref{Fig:K3cotD1}. A dashed curve with a band indicates the final fit result with 1 standard deviation, 
which is obtained from renewed fitting on three new data together with four data.}
\label{Fig:K1cotD0_Full}
\end{figure}
%

Combined with this $P$-wave information, we also evaluate $k{\rm{cot}}\delta_{0}$ at very low energies 
using the data calculated under other twisted boundary conditions with a twist angle vector $\vec{\theta} = (0, 0, \theta)$, through the finite size formula
%
\beq
{\rm{cot}}\delta_{0}(k) &=& {\cal M}^{[001]}_{SS}(q) + \frac{ | {\cal M}_{SP}^{[001]}(q) |^{2} }{ {\rm{cot}}\delta_{1}(k) - {\cal M}^{[001]}_{PP}(q) }\cr
&=&
w_{00}(q)+\frac{3w_{10}^2(q)}{\cot \delta_1(k)- w_{00}(q)-2w_{20}(q)}.
\label{Eq:FVFormula_100}
\nonumber\\
\eeq
Here we use three different angles, $\theta=1.44$, 2.16 and 2.88 (rad), to choose $k^2$ points that
range from 0 to $\left(\frac{\pi}{L}\right)^2$ in lattice units.

Figure \ref{Fig:K1cotD0_Full} shows the results of $k {\rm{cot}}\delta_{0}$. We have carried out 
correlated $\chi^2$ fits on all seven data displayed in Fig.~\ref{Fig:K1cotD0_Full}
by using the form of the effective-range expansion:
\beq
k{\rm{cot}}\delta_{0}(k) = \frac{1}{a_{0}} + \frac{1}{2} r_{0} k^{2} + \sum_{n=2}^{N}v_{n}k^{2n}
\label{Eq:EffRangeFit}
\eeq 
up to $N=4$, which corresponds to the order of $k^8$. 
We recall that the effect of $D$-wave contributions is, strictly speaking, not negligible at the order of $k^4$; 
the coefficient $v_2$ in the $k^{4}$ term, which corresponds to the scattering volume $pr_0^3$, may suffer from 
systematic uncertainties due to such unknown effects of higher partial-wave mixing. 
However, we will quote the coefficient $v_2$ as a reference value of the scattering volume 
as well as the scattering length $a_0$ and the effective range $r_0$, which are safely 
read off from the coefficients of the $k^0$ and $k^2$ terms later.

The stability of the fit results has been tested against 
either the number of fitted data points or the number of polynomial terms for
a given order $N$ defined by Eq.~(\ref{Eq:EffRangeFit}).
The best fit is drawn to fit all seven data points using the fitting form (\ref{Eq:EffRangeFit}) with $N=4$ in Fig.~\ref{Fig:K1cotD0_Full}.
We then obtain the following results:
\beq
a_{0} 
&=& 0.242 \pm 0.041 \ \ [{\rm{fm}}], \nonumber \\
r_{0}
&=& 4.712 \pm 1.727 \ \ [{\rm{fm}}], \nonumber \\
pr_{0}^{3}
&=& 2.50 \pm 1.64 \ \ [{\rm{fm}}^{3}],\nonumber 
\eeq
where the quoted errors represent only the statistical errors given by the jackknife analysis.
The effective range $r_{0}$ receives a rather large error.
This is because the size of the effective range is much larger than that of the scattering length, $r_0\gg a_0$.
In this particular case, the coefficient of the linear term in momentum space is influenced even 
by small statistical fluctuations at lower energy data points.
More statistics are clearly needed to reduce the statistical error on the effective range.
But, we would like to stress here that our approach shows feasibility for determining the effective range
as well as the scattering length. 

\begin{figure*}
\begin{tabular}{cc}
\begin{minipage}{0.5\hsize}
\begin{center}
\includegraphics[width=0.9 \textwidth]{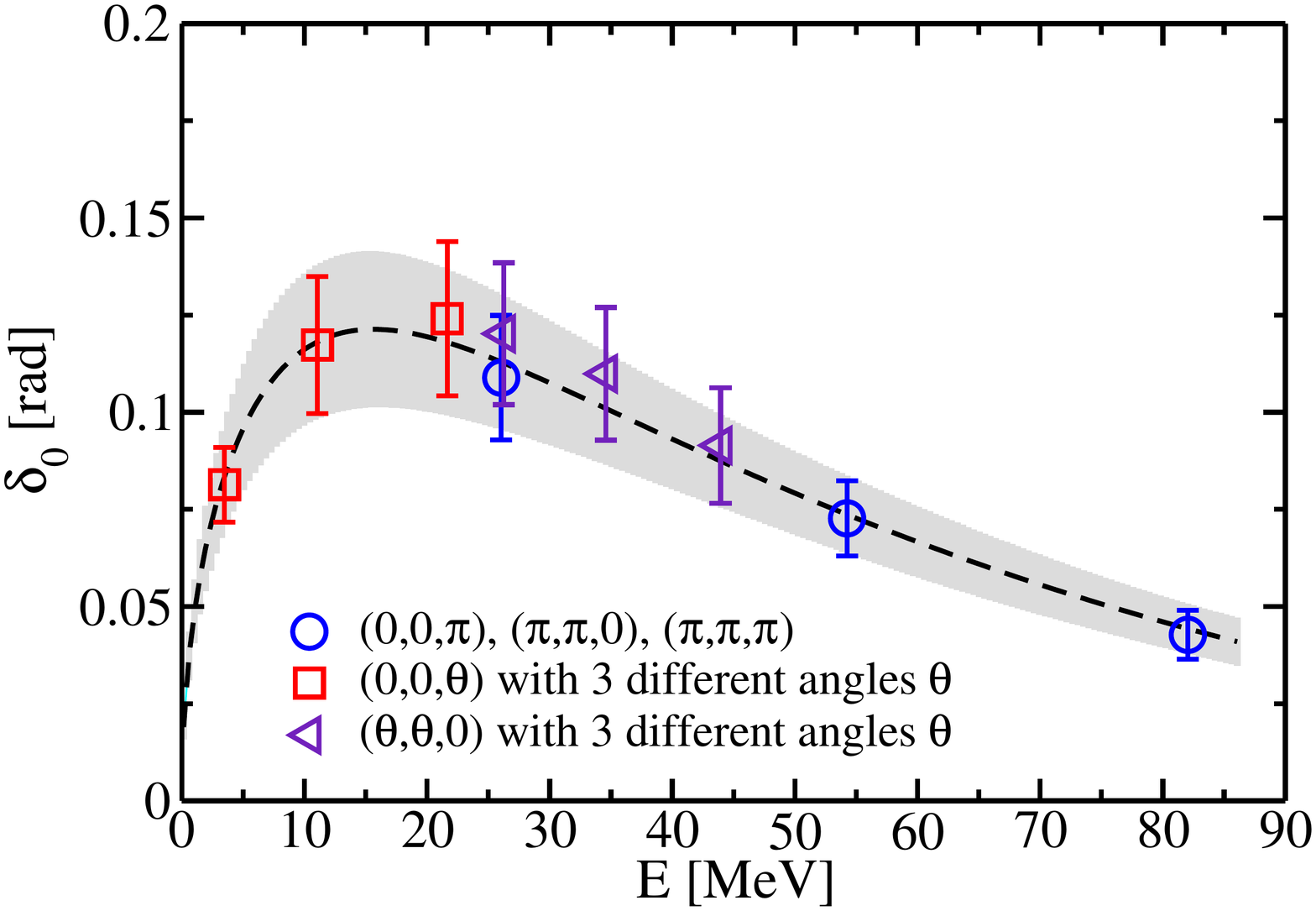}
\end{center}
\end{minipage}
\begin{minipage}{0.5\hsize}
\begin{center}
\includegraphics[width=0.9 \textwidth]{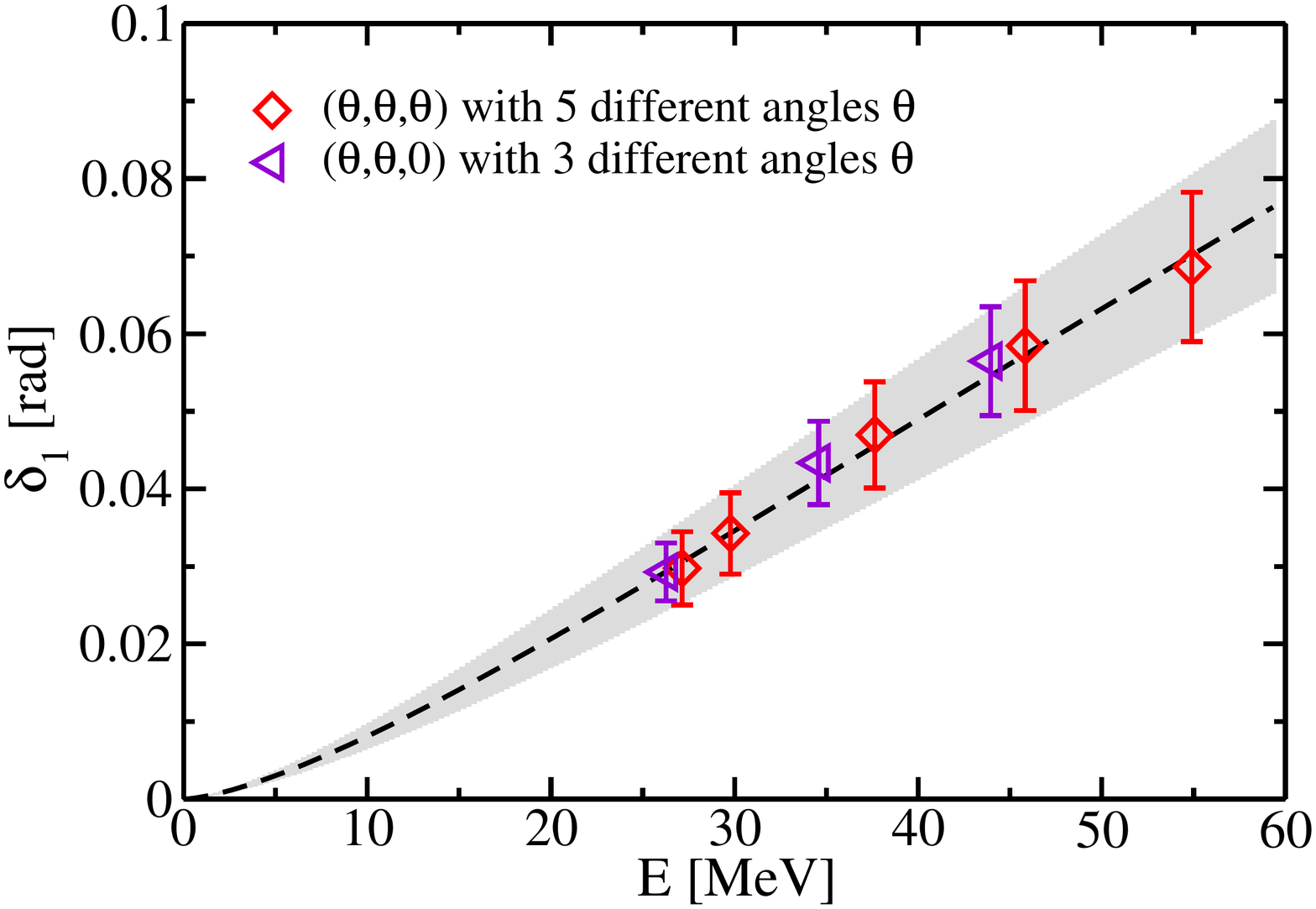}
\end{center}
\end{minipage}
\end{tabular}
\caption{$S$-wave (left panel) and $P$-wave (right panel) scattering 
phase shifts as a function of 
the two-particle energy $E$, which is measured from the $\JPP$ threshold.
In each panel, a dashed curve represents the fit result guided by the effective range expansion applied 
to all data points except for open left-triangle symbols, which are additionally calculated with the other
twist angle vector $\vec{\theta}=(\theta,\theta,0)$.}
\label{Fig:PhaseShift_Full}
\end{figure*}

Figure~\ref{Fig:PhaseShift_Full} shows $S$-wave (left panel) and $P$-wave (right panel) scattering 
phase shifts for the $J/\psi$-$\phi$ system. The phase shifts observed here are 
positive reflecting the attractive interaction between the $J/\psi$ and $\phi$ states 
in both channels. 
In each panel, a dashed curve represents the fit result guided by the effective range expansion applied 
to all data points except for open left-triangle symbols, which are additionally calculated with the other
twist angle vector $\vec{\theta}=(\theta,\theta,0)$. The fit result of the $P$-wave phase shift is used as prior information 
in the $S$-wave calculation for new data, and vice versa. New data points in both panels show an excellent 
agreement with the fit curves. This means that the results obtained from our data analysis developed here
pass the consistency test. 

%
%
%

Our results exhibit typical behaviors of weak attraction (small scattering length and small scattering volume) 
in $S$-wave and $P$-wave phase shifts at low energies. Unfortunately, there is no resonance structure associated 
with the $Y(4140)$ resonance against what we expected. Although the CDF experiment has reported evidence 
for a new narrow resonance, $Y(4140)$ around a few 10 MeV above the $\JPP$ threshold, 
it seems that our results are consistent with null results in the Belle and LHCb experiments.

Next, we give some remarks on how large the effect of the higher partial-wave mixing remains at low energies.
Although the lower partial wave becomes dominant at low energies due to $\delta_l(p) \propto a_l p^{2l+1}$,
the $P$-wave mixing induced by the usage of the partially twisted boundary conditions 
should be suppressed in the limit of $\theta \rightarrow 0$ or $L\rightarrow \infty$. 
If we assume $|{\cal M}^{\vec{\theta}}_{SP}|\approx 0$ at low energies, the finite size formula
then reduces to the original L\"uscher's formula
\beq
\cot \delta_0(k)=
w_{00}(q)=\frac{1}{\pi^{3/2}q}Z_{00}^{\vec{\theta}}(1;q^2)
\label{Eq:ReducedFS}
\eeq
with the twist angle vector $\vec{\theta}$.

\begin{figure}[t]
\begin{minipage}{\hsize}
\begin{center}
\includegraphics[width=0.9 \textwidth]{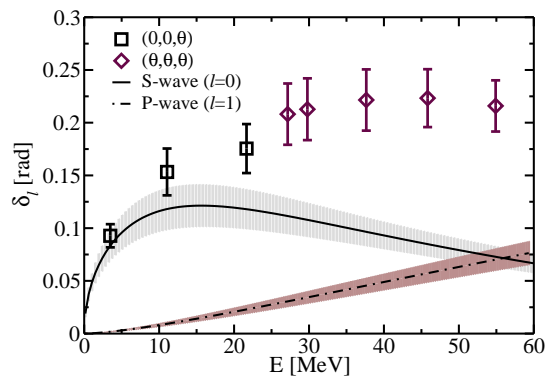}
\end{center}
\end{minipage}
\caption{Scattering phase shifts as a function of the two-particle energy $E$. 
Open symbols are calculated through the approximated formula (\ref{Eq:ReducedFS}), where
the $P$-wave mixing is ignored, for data taken with two types of twist angle vectors, $\vec{\theta}=(0,0,\theta)$ and $(\theta,\theta,\theta)$.
Solid and dashed curves correspond to the $S$-wave and $P$-wave phase shifts
given in the full analysis, as described in Fig.~\ref{Fig:PhaseShift_Full}.}
\label{Fig:PhaseShift_nomix}
\end{figure}

Figure~\ref{Fig:PhaseShift_nomix} exposes the size of the systematic error in the analysis 
where the $P$-wave mixing is ignored.
Open symbols are calculated through the approximated formula (\ref{Eq:ReducedFS})
for data taken with two types of twist angle vectors, $\vec{\theta}=(0,0,\theta)$ and $(\theta,\theta,\theta)$.
In Fig.~\ref{Fig:PhaseShift_nomix}, a solid curve corresponds to the $S$-wave phase shift in the
full analysis as described previously, while a dashed curve represents the $P$-wave phase shift
in the same analysis. When the $P$-wave phase shift is smaller enough than the $S$-wave phase shift,
$\delta_0 \gg \delta_1$, the difference between the results from the full and approximated analyses gets smaller,
especially near the threshold, as we expected. However, the validity region of the approximation, such that $|{\cal M}^{\vec{\theta}}_{SP}|=0$, 
is limited in the vicinity of the threshold even for such weakly interacting systems, where both the scattering length and 
the scattering volume are observed to be small. 

From this observation, our ignorance of the higher partial wave ($l\ge 2$) in the full analysis
would be acceptable, at least up to a few 10 MeV above the threshold, where the $P$-wave mixing is safely ignored.
We therefore expect that our null result for the $Y(4140)$ resonance in this vicinity in both $S$- and $P$-wave channels 
remains unchanged even if the $D$-wave mixing is taken into account in the finite size formula. Nevertheless, we plan to develop our analysis, 
with the finite size formula extended to treat the higher-wave mixing up to $l=2$, in a separate publication.
As such, the formula has already been considered for the case of general moving-frame computation~\cite{Gockeler:2012yj}.

Finally, we make a comment about the usage of the partially twisted boundary conditions, which
are imposed only on valence quarks and then may induce some systematic error due to the 
difference of boundary conditions between dynamical and valence quarks. This is only an issue for
the strange quark in the $\JPP$ system, since the charm quark is treated as a quenched approximation.
The authors of Ref.~\cite{Sachrajda:2004mi} show that the finite volume correction due to the partially twisted 
boundary conditions is exponentially suppressed as the spatial extent $L$ increases. The coefficient of exponential-type
finite size corrections may be characterized by a mass of the lightest state mediated between two-particles.
In the $J/\psi$-$\phi$ system, a single meson exchange such as one pion exchange is forbidden, unlike 
a typical hadron-hadron interaction, since the $J/\psi$ and $\phi$ states do not carry any isospin
and also do not share the same quark flavor. 
Therefore, instead of the complex nature originating from quark exchanges, multigluon exchanges become 
dominant in the $\JPP$ system. An effective description for the interaction between the $J/\psi$ and $\phi$
may be given by the soft Pomeron exchange, which carries the quantum number of the vacuum~\cite{Landshoff:1986yj}.  
In this sense, the difference of boundaries between dynamical quarks and valence quarks in this study 
would be negligibly small since the mass of the Pomeron is typically of the order of $500$$-$$600$ MeV, which is, phenomenologically, a
good agreement with what is observed through the charmonium-nucleon potential~\cite{{Kawanai:2010ru},{Kawanai:2010ev}}.

%
%
%

\section{Summary}

We have considered how to develop the L\"uscher finite size approach under the partially twisted boundary condition.
The twisted boundary condition allows us to treat any small momentum on the lattice through the variation of the twist angle, 
continuously. On the other hand, when the twisted boundary condition is introduced,
the cubic symmetry is broken down to some subgroup symmetry, where the center of inversion symmetry is unfortunately lost. 
Accordingly, even-$l$ and odd-$l$ partial waves are inevitably mixed together in the extended L\"uscher finite size formula.

We propose to take advantage of these unique properties of the twisted boundary condition
in order to extract both the $S$-wave and $P$-wave scattering phase shifts from the two-particle energy with the total
momentum $|\vec{P}|=0$ calculated in a single finite box with various types of partially twisted boundary conditions.
We then demonstrate the feasibility of a new approach to examine the existence of near-threshold and narrow resonance states, 
whose characteristics are observed in many of the newly discovered charmonium-like $XYZ$ mesons.

As an example, we choose low-energy $\JPP$ scatterings to search for the $Y(4140)$ resonance, which is observed around 
a few 10 MeV above the $\JPP$ threshold by the CDF experiment at Fermilab 
through the $B^+ \rightarrow J/\psi\phi K^+$ decay process from $p\bar{p}$ collisions.
Our simulations are performed in 2+1 flavor dynamical lattice QCD using the PACS-CS gauge configurations
at the lightest pion mass, $m_{\pi}=156$ MeV, with a relativistic heavy-quark action for the charm quark. 
We successfully obtain the $P$-wave phase shift of the $\JPP$ scattering as well as the $S$-wave phase shift near the threshold.

Our results exhibit typical behaviors of weak attraction (small scattering length and small scattering volume) 
in both $S$-wave and $P$-wave channels at low energies. There is no resonance structure associated with the $Y(4140)$ 
resonance, contrary to what we expected. Therefore, our results are consistent with null results in the Belle and LHCb experiments.
Instead of finding near-threshold and narrow resonance states, 
we succeed in extracting model-independent information of the low-energy $\JPP$ interaction,
such as the scattering length $a_0=0.242\pm0.041$ fm and the effective range $r_0=4.71\pm1.73$ 
fm for the $S$-wave and the scattering volume 
$a_1= 0.035\pm 0.008$ ${\rm fm}^3$ for the $P$-wave, within our new approach. 
We plan to apply this new method to nucleon-nucleon scatterings and also $D\textit{\rm -}K$ and $D^{*}\textit{\rm -}K$ 
scatterings. The former is a key target in the original proposal of twisted boundary conditions~\cite{Bedaque:2004kc}, while
the latter would give some insight into the structure of the $D_{s0}(2317)$ and $D_{s1}(2460)$ resonances.


\begin{acknowledgments}
We would like to thank T. Hatsuda and T. Kawanai for fruitful discussions. 
This work is supported by MEXT Grant-in-Aid for Scientific Research on Innovative Areas
(No.2004:20105003) and JSPS Grants-in-Aid for Scientific Research (C)
 (No.~23540284). Numerical calculations reported here were carried out on the T2K supercomputer at ITC, 
 University of Tokyo and also at CCS, University of Tsukuba.
\end{acknowledgments}

%


\begin{thebibliography}{99}




\bibitem{Luscher:1985dn}
  M.~L\"uscher,
  Commun.\ Math.\ Phys.\  {\bf 105}, 153 (1986).
  
\bibitem{Luscher:1990ux}
  M.~L\"uscher,
  Nucl.\ Phys.\ B {\bf 354}, 531 (1991).


\bibitem{Beane:2008dv} 
  S.~R.~Beane, K.~Orginos and M.~J.~Savage,
  Int.\ J.\ Mod.\ Phys.\ E {\bf 17}, 1157 (2008)
  and references therein.
  
\bibitem{foot1}
Signatures of 
$S$-wave bound state formation in finite volume can also be discussed 
on the basis of L\"uscher's method~\cite{{Beane:2003da},{Sasaki:2006jn}}.
 
\bibitem{Beane:2003da} 
  S.~R.~Beane, P.~F.~Bedaque, A.~Parreno and M.~J.~Savage,
  Phys.\ Lett.\ B {\bf 585}, 106 (2004).
   
\bibitem{Sasaki:2006jn} 
  S.~Sasaki and T.~Yamazaki,
  Phys.\ Rev.\ D {\bf 74}, 114507 (2006).
 

\bibitem{Rummukainen:1995vs}
  K.~Rummukainen and S.~A.~Gottlieb,
  Nucl.\ Phys.\  B {\bf 450}, 397 (1995).

\bibitem{Christ:2005gi} 
  N.~H.~Christ, C.~Kim and T.~Yamazaki,
  Phys.\ Rev.\ D {\bf 72}, 114506 (2005).

\bibitem{Kim:2005gf} 
  C.~h.~Kim, C.~T.~Sachrajda and S.~R.~Sharpe,
  Nucl.\ Phys.\ B {\bf 727}, 218 (2005).


  
\bibitem{Feng:2004ua}
  X.~Feng, X.~Li and C.~Liu,
  Phys.\ Rev.\  D {\bf 70}, 014505 (2004).
 

\bibitem{Li:2007ey}
  X.~Li {\it et al.}  [CLQCD Collaboration],
  JHEP {\bf 0706}, 053 (2007).
  
  
\bibitem{Bedaque:2004kc}
  P.~F.~Bedaque,
  Phys.\ Lett.\  B {\bf 593}, 82 (2004).


\bibitem{Newton:1982qc}
  R.~G.~Newton, 
  ``Scattering Theory of Waves and Particles'', 2nd ed. (Springer, New York, 1982).


\bibitem{Ozaki}
 S.~Ozaki and S.~Sasaki,
 PoS {\bf LATTICE2012}, 160 (2012).
 
   
  
\bibitem{Brambilla:2010cs} 
  N.~Brambilla, S.~Eidelman, B.~K.~Heltsley, R.~Vogt, G.~T.~Bodwin, E.~Eichten, A.~D.~Frawley and A.~B.~Meyer {\it et al.},
  Eur.\ Phys.\ J.\ C {\bf 71}, 1534 (2011).


  \bibitem{Godfrey:2008nc}
  S.~Godfrey and S.~L.~Olsen,
  Ann.\ Rev.\ Nucl.\ Part.\ Sci.\  {\bf 58}, 51 (2008).


\bibitem{Belle:2011aa} 
  A.~Bondar {\it et al.}  [Belle Collaboration],
  Phys.\ Rev.\ Lett.\  {\bf 108}, 122001 (2012).
  
\bibitem{Collaboration:2011gja} 
  I.~Adachi [Belle Collaboration],
  arXiv:1105.4583 [hep-ex].


\bibitem{Abe:2007tk} 
  K.~F.~Chen {\it et al.}  [Belle Collaboration],
  Phys.\ Rev.\ Lett.\  {\bf 100}, 112001 (2008).

\bibitem{Chen:2008xia} 
  K.~-F.~Chen {\it et al.}  [Belle Collaboration],
  Phys.\ Rev.\ D {\bf 82}, 091106 (2010).
 
  
  
\bibitem{Aaltonen:2009tz}
  T.~Aaltonen {\it et al.}  [CDF Collaboration],
  Phys.\ Rev.\ Lett.\  {\bf 102}, 242002 (2009).
  
\bibitem{Aaltonen:2011at}
  T.~Aaltonen {\it et al.}  [The CDF Collaboration],
  arXiv:1101.6058 [hep-ex].
  
\bibitem{Shen:2009vs}
  C.~P.~Shen {\it et al.}  [Belle Collaboration],
  Phys.\ Rev.\ Lett.\  {\bf 104}, 112004 (2010).
  
\bibitem{Aaij:2012pz} 
  R. Aaij {\it et al.}  [LHCb Collaboration],
  Phys.\ Rev.\ D {\bf 85}, 091103 (2012).
  
  


\bibitem{Sachrajda:2004mi} 
  C.~T.~Sachrajda and G.~Villadoro,
  Phys.\ Lett.\ B {\bf 609}, 73 (2005).


  
\bibitem{deDivitiis:2004kq}
  G.~M.~de Divitiis, R.~Petronzio and N.~Tantalo,
  Phys.\ Lett.\  B {\bf 595}, 408 (2004).
  
\bibitem{Boyle:2008yd}
  P.~A.~Boyle {\it et al.},
  JHEP {\bf 0807}, 112 (2008).
  
\bibitem{Kim:2010sd}
  C.~H.~Kim and C.~T.~Sachrajda,
  Phys.\ Rev.\  D {\bf 81}, 114506 (2010).


\bibitem{Davoudi:2011md} 
  Z.~Davoudi and M.~J.~Savage,
  Phys.\ Rev.\ D {\bf 84}, 114502 (2011).
  
\bibitem{Fu:2011xz}
  Z.~Fu,
  Phys.\ Rev.\  D {\bf 85}, 014506 (2012).
  
\bibitem{Leskovec:2012gb} 
  L.~Leskovec and S.~Prelovsek,
  Phys.\ Rev.\ D {\bf 85}, 114507 (2012).
  
\bibitem{Doring:2012eu} 
  M.~D\"oring, U.~G.~Meissner, E.~Oset and A.~Rusetsky,
  Eur.\ Phys.\ J.\ A {\bf 48}, 114 (2012).

\bibitem{Gockeler:2012yj} 
  M.~M. G\"ockeler, R.~Horsley, M.~Lage, U.~-G.~Meissner, P.~E.~L.~Rakow, A.~Rusetsky, G.~Schierholz and J.~M.~Zanotti,
  Phys.\ Rev.\ D {\bf 86}, 094513 (2012).
  
\bibitem{Kawanai:2010ru} 
  T.~Kawanai and S.~Sasaki,
  PoS LATTICE {\bf 2010}, 156 (2010)
  [arXiv:1011.1322 [hep-lat]].
  
\bibitem{foot2}
The Lorentz transformation 
is characterized by the velocity $\vec{v}=\vec{P}/E$, where $E$ and $\vec{P}$ are the total energy 
and the total momentum, respectively, and also the Lorentz boost factor $\gamma=1/\sqrt{1-\vec{v}^2}$.
  
\bibitem{Yamazaki:2004qb} 
  T.~Yamazaki {\it et al.}  [CP-PACS Collaboration],
  Phys.\ Rev.\ D {\bf 70}, 074513 (2004).
  
  

\bibitem{Aoki:2008sm}
  S.~Aoki {\it et al.}  [PACS-CS Collaboration],
  Phys.\ Rev.\  D {\bf 79}, 034503 (2009).

  \bibitem{ILDG}
  http://www.usqcd.org/ildg (ILDG) and
  http://www.jldg.org (JLDG).

  \bibitem{ElKhadra:1996mp}
  A.~X.~El-Khadra {\it et al.}, 
  Phys.\ Rev.\  D {\bf 55}, 3933 (1997).

  \bibitem{Aoki:2001ra}
  S.~Aoki, Y.~Kuramashi and S.~I.~Tominaga,
  Prog.\ Theor.\ Phys.\  {\bf 109}, 383 (2003).
  
  \bibitem{Kayaba:2006cg}
  Y.~Kayaba {\it et al.}  [CP-PACS Collaboration],
  JHEP {\bf 0702}, 019 (2007).

   
  \bibitem{Kawanai:2011jt} 
  T.~Kawanai and S.~Sasaki,
  Phys.\ Rev.\ D {\bf 85}, 091503 (2012).




  
\bibitem{McNeile:2004wu} 
  C.~McNeile {\it et al.}  [UKQCD Collaboration],
  Phys.\ Rev.\ D {\bf 70}, 034506 (2004).
  
\bibitem{deForcrand:2004ia} 
  P.~de Forcrand {\it et al.}  [QCD-TARO Collaboration],
  JHEP {\bf 0408}, 004 (2004).


\bibitem{Moore:2005dw} 
  D.~C.~Moore and G.~T.~Fleming,
  Phys.\ Rev.\ D {\bf 73}, 014504 (2006)
  [Erratum-ibid.\ D {\bf 74}, 079905 (2006)].

\bibitem{Foley:2012wb} 
  J.~Foley, J.~Bulava, Y.~-C.~Jhang, K.~J.~Juge, D.~Lenkner, C.~Morningstar and C.~H.~Wong,
  PoS LATTICE {\bf 2011}, 120 (2011)
  [arXiv:1205.4223 [hep-lat]].


\bibitem{foot3}
The index $g$ (gerade) or $u$ (ungerade) denotes the existence of an inversion center and indicates even or odd parity.
  
\bibitem{Yokokawa:2006td} 
  K.~Yokokawa, S.~Sasaki, T.~Hatsuda and A.~Hayashigaki,
  Phys.\ Rev.\ D {\bf 74}, 034504 (2006).


\bibitem{foot4}
For the case of $\vec{\theta}=(0,0,0)$, where $G$ is given by $O_h$, 
the condition of $R \vec{p} = \vec{p}$ is automatically satisfied since $\vec{p}=0$.

\bibitem{foot5}
The quadratic coefficient in the $k^{4}$ term should be unphysical, since the effect of higher partial-wave ($D$-wave) contributions is not negligible at the order of $k^4$.


\bibitem{Landshoff:1986yj} 
  P.~V.~Landshoff and O.~Nachtmann,
  Z.\ Phys.\ C {\bf 35}, 405 (1987).
    
\bibitem{Kawanai:2010ev} 
  T.~Kawanai and S.~Sasaki,
  Phys.\ Rev.\ D {\bf 82}, 091501 (2010).
  

      
\end{thebibliography}
\end{document}